\documentclass[a4paper, aps, prd,nofootinbib, notitlepage, preprintnumbers , 10pt]{revtex4-1}% twocolumn
\usepackage[T1]{fontenc}
\usepackage[latin1]{inputenc}
\usepackage[english]{babel} 
\usepackage{amsmath}
\usepackage{amsfonts}
\usepackage{mathtools}
\usepackage{bbm}
\usepackage{accents}
\usepackage{dsfont}
\usepackage{mathrsfs}
\usepackage[dvips]{graphicx}
\usepackage{color}
\usepackage{graphicx}
\newcommand{\qq}[1]{``#1''} %Anfuehrungszeichen
\usepackage{hyperref}
\newcommand{\utilde}[1]{\undertilde{#1}}
\newcommand{\di}{\mathrm{d}} %Differential
\newcommand{\ou}[3]{{#1}{}^{#2}{}_{#3}} %Indexstellung
\newcommand{\uo}[3]{{#1}{}_{#2}{}^{#3}} %Indexstellung
\newcommand{\I}{\mathrm{i}} %imaginaere Einheit
\newcommand{\E}{\mathrm{e}} %Euler Zahl
\newcommand{\ellp}{{\ell_{\mathrm{P}}}} %Planck Laenge
\newcommand{\CC}{\mathrm{cc.}} % komplex konjugiertes
\newcommand{\eref}[1]{(\ref{#1})}

\newcommand{\C}{\mathbb{C}}

\newcommand{\R}{\mathbb{R}}
\newcommand{\Z}{\mathbb{Z}}
\newcommand{\T}{\mathbb{T}}
\newcommand{\su}{{\mathfrak{su}}}

\renewcommand{\Im}{\mathfrak{Im}}

\makeatletter
\renewcommand{\subparagraph}{%
  \@startsection{paragraph}{4}%
  {\z@}{3.25ex \@plus 1ex \@minus .2ex}{-1em}%
  {\normalfont\normalsize\bfseries}%
}
\makeatother

\begin{document}
\title{New action for simplicial gravity in four dimensions}
\author{Wolfgang M. Wieland}
\email{wieland@gravity.psu.edu}
\affiliation{
Institute for Gravitation and the Cosmos \& Physics Department\\
104 Davey Laboratory, PMB \#070\\
University Park, PA 16802-6300, U. S. A.
}

\preprint{IGC-14/6-3}
\date{June 2014}
\begin{abstract}\noindent
%Spinors have a wide range of applications, from quantum mechanics to particle physics, quantum information and general relativity. In this paper, we will argue that they are useful also for discretized gravity, and develop a proposal for  
We develop a proposal for a theory of simplicial gravity with spinors as the fundamental configuration variables. The underlying action describes a mechanical system with finitely many degrees of freedom, the system has a Hamiltonian and local gauge symmetries. We will close with some comments on the resulting quantum theory, and explain the relation to loop quantum gravity and twisted geometries. The paper appears in parallel with an article by Cortês and Smolin, who study the relevance of the model for energetic causal sets and various other approaches to quantum gravity.
\end{abstract}
\maketitle
  
%-------------------------------------------------------------------------------%
\section{Introduction}\noindent
%-------------------------------------------------------------------------------%
%\alert{Da fehlt no allerhand, Idee: Regges Diskretisierung der allgemeinen Relativit\"atstheorie ist eine koordinatenfreie -- also eichinvariante -- Formulierung der einsteinschen Theorie auf einem Gitter. Alle Materiefreiheitsgrade koppeln aber gerade vermittels eichkovarianter Ausdr\"ucke an die Gravitation. Damit sind wir schon bei der Fragestellung dieses Artikels angekommen: Wie l\"asst sich eine 
%Regge--gauge invariance--gauge dependence because of matter.}
%How was Regge calculus \cite{Reggecalc} originally motivated? 
%Regge calculus was originally motivated as a coordinate-free formulation of general relativity on a lattice, as an attempt to discretize general relativity using only gauge invariant variable. Regge calculus is trivially invariant under local Lorentz transformations, the question whether it is also invariant under general coordinate transformations should not bother us here. The point we would like to make is this: \cite{carlogauge}
Regge calculus \cite{Reggecalc} was originally motivated as a coordinate-free formulation of general relativity on a lattice, as an attempt to discretize general relativity with only gauge-invariant quantities. Yet, gauge symmetries are not just a  mathematical redundancy, gauge symmetries are crucial \cite{carlogauge} for the coupling between different physical systems: A fermion couples to gravity through both the tetrad and an \emph{a priori} independent spin connection, through \emph{gauge-variant} quantities, or \emph{partial observables} in the terminology of \cite{rovelli}. 

If we then rather sought for a version of \emph{first-order} gravity on a lattice, then there would already be a comprehensive library at hand, see \cite{Reggebarrett, Barrett:1994nn,ReggeDittrichSimone,Bahr:2009qd,PhysRevD.82.064026,Dittrich:2008ar,Khatsymovsky2002321,Drummond1986125, Obukhov:2003aa, Caselle:1989hd, Gronwald:1995mb, Pereira:2002ff} and especially \cite{Reggebarrett,ReggeDittrichSimone,PhysRevD.82.064026}. Here, we are, however, seeking for an answer to a subtler question: Is there a \emph{Hamiltonian} formalism  for discretized gravity in terms of first-order tetrad-connection variables available? This is a difficult question, because there is a conceptual tension: A Hamiltonian always generates a differential equation, it generates a Hamiltonian flow, while, on the other hand, discretized theories are typically governed by difference equations instead \cite{Dittrich:2013jaa}.

This article develops a proposal resolving the tension. Following the Pleba\'{n}ski principle, we start with the topological BF action \cite{Horowitz:1989ng}. We introduce a simplicial decomposition of the four-dimensional spacetime manifold, and discretize the action. %  over the elementary two-dimensional surfaces. 
This leads us to a sum over the two-dimensional simplicial faces. %one-dimensional integrals,---one for each face in the discretization. % These integrals define an action for a one-dimensional mechanical system. 
Every face contributes a one-dimensional integral over its bounding edges, thus turning the topological action into an integral over the entire system of edges---into an an action over a \emph{one-dimensional branched manifold}.
% Now, every face is bounded by several edges, and each of these edges belongs to many faces. The total action for the discretized theory is, therefore, a one-dimensional integral over an entire network of edges. The edges do not have open ends, they meet at vertices, hence the action is an integral over a \emph{branched one-dimensional manifold}. 

%The $t$-coordinates parametrizing this one-dimensional manifold provide us with a natural notion of time. We perform a Hamiltonian analysis and find a phase space, local gauge symmetries and a Hamiltonian. %These $t$-variables do however not measure time or duration, they are nothing but coordinates, and do not exist globally. %The resulting equations of motion correspond to a discretization of the equations of motion for the continuum theory: We could either start with the continuous action, derive the equations of motion for the continuum  theory, and then discretize a solution over the elementary edges; or we do the opposite, and first discretize the action in order to derive the equations of motion for the discretized theory: In both cases we would get the same result, thus proving the internal consistency of the procedure.

The next step is to study the simplicity constraints and add them to the discretized action. The simplicity constraints break the topological shift symmetries of BF theory \cite{Freidel:2002dw}, and impose that the BF configuration variables are compatible with the existence of a  metric. In the continuum, the simplicity constraints reduce the BF action to an action for general relativity. We assume that this is also true in the discrete theory, and add these conditions in the form of the linear simplicity constraints \cite{flppdspinfoam, LQGvertexfinite}, which introduce an additional element to the theory---the volume-weighted time normals of the elementary tetrahedra. We will argue that a robust theory should treat these time-normals as dynamical variables. We will propose such a theory, and prove an intriguing correspondence: The entire simplicial complex represents a system of free particles propagating in a locally flat auxiliary spacetime, with every tetrahedron representing one of those particles, the volume-weighted time-normals representing the particles' four-momenta, %the three-volume of every single tetrahedron playing the role of the particles' rest mass.
and the entire discretized action turning into an integral over the worldlines of those auxiliary particles.

Of the following sections, the first deals with the derivation of the action, while the second studies the resulting equations of motion and the Hamiltonian formulation of the theory. The system has a phase space, local gauge symmetries and a Hamiltonian. The underlying phase space is the space of $SL(2,\C)$ holonomy-flux variables in the spinorial representation \cite{twistconslor,komplexspinors,twistintegrals,twist}. Spinors do not change the physics of the theory, yet they are useful for us because they embed the nonlinear phase space $T^\ast SL(2,\C)$ of holonomy-flux variables into a linear space with global Darboux coordinates.  

The Hamiltonian is a sum over constraints, which are of both first and second class. We show that all constraints are preserved under the physical time evolution, hence the model passes its first consistency check. Another test for the model is whether the solutions of the equations of motion have a geometric interpretation, whether there is a chance to describe a physical spacetime. Here we only have a partial result: We will show that a generic solution of the equations of motion represents a so-called \emph{twisted geometry}. Twisted geometries \cite{freidelsimotwist, twist, twistedconn} are a generalization of Regge geometries. In a twisted geometry every tetrahedron has a well-defined volume, and every triangle has a unique area, yet there are no unique length variables: Given a collection of tetrahedra sharing a triangle, the shape of the triangle is different depending on whether we compute it from the metric
in one tetrahedron or the other. Our last consistency check concerns the deficit angle around a triangle. We find that for a general solution of the equations of motion the Lorentz angles between adjacent tetrahedra do not sum up to zero, hence there is curvature in a face. 

The paper appears in parallel with an article \cite{CortezLee} by Cortês and Smolin, who study the relation to energetic causal sets, and further explain the relevance of the model for other approaches to quantum gravity, including loop quantum gravity \cite{status, thiemann, zakolec} and relative locality \cite{relloc}.

%-------------------------------------------------------------------------------%
\section{Construction of the action}\label{secI}\noindent
%-------------------------------------------------------------------------------%
This section is already the main part of the paper. Here will we develop our proposal for a new action for simplicial gravity in terms of first-order area-connection variables. We will derive this action by translating the Pleba\'nski formalism \cite{Krasnov:2009aa, Krasnov:2011aa}---general relativity written as a constrained topological theory---to the simplicial setting. Indeed, the topological BF action\footnote{Let us briefly fix our conventions:  $\ou{\Sigma}{\alpha}{\beta}$ is an $\mathfrak{so}(1,3)$-valued two-form in the four-dimensional spacetime manifold $M$, $\ou{A}{\alpha}{\beta}$ is an $SO(1,3)$ connection, with $\ou{F}{\alpha}{\beta}=\di\ou{A}{\alpha}{\beta}+\ou{A}{\alpha}{\mu}\wedge\ou{A}{\mu}{\beta}$ denoting its curvature; $\ellp^2=8\pi\hbar G/c^3$ is the Planck area, hence $\ou{\Sigma}{\alpha}{\beta}$ has dimensions of an area, while $\beta>0$ is the Barbero--Immirzi parameter. The flat Minkowski metric $\eta_{\alpha\beta}$ of signature $(-+++)$ moves all internal Lorentz indices $\alpha,\beta,\gamma,\dots$, and $\ast\Sigma_{\alpha\beta}=\tfrac{1}{2}\uo{\epsilon}{\alpha\beta}{\mu\nu}\Sigma_{\mu\nu}$ defines the Hodge dual in internal space.}
\begin{equation}
 S_{\text{BF}}[\Sigma,A]=\frac{\hbar}{2\ellp^2}\int_M\big(\ast \Sigma_{\alpha\beta}-\beta^{-1}\Sigma_{\alpha\beta}\big)\wedge F^{\alpha\beta}[A]\equiv\int_{M}\Pi_{\alpha\beta}\wedge F^{\alpha\beta},\label{BFactn}
\end{equation}
determines the symplectic structure\footnote{The pull-back of $\ou{\Pi}{\alpha}{\beta}$ and $\ou{A}{\alpha}{\beta}$ to a spatial slice defines a pair of canonically conjugate fields.} of the theory, general relativity is a consequence of the \emph{simplicity constraints}
\begin{equation}
\Sigma^{\alpha\beta}\wedge\Sigma^{\mu\nu}\propto\epsilon^{\alpha\beta\mu\nu}\label{simplcons}
\end{equation}
added to the action. These constraints imply the geometricity of the Pleba\'nski two-form $\Sigma_{\alpha\beta}$, that is the existence of a tetrad $e_\alpha$ such that $\Sigma_{\alpha\beta}$ is \emph{simple}\footnote{There is also the twisted solution $\Sigma_{\alpha\beta}=\ast(e_\alpha\wedge e_\beta)=\frac{1}{2}\uo{\epsilon}{\alpha\beta}{\mu\nu}e_\mu\wedge e_\nu$. This solution would again lead to an action for general relativity, with the Barbero--Immirzi parameter $\beta$ and Newton's  constant $G$ turning into $-\beta^{-1}$ and $\beta G$ respectively. An advantage of the \emph{linear} simplicity constraints, which we will introduce in a second, is that they omit this solution.}: $\Sigma_{\alpha\beta}=e_\alpha\wedge e_\beta$. Imposing the simplicity constraints reduces therefore the topological action \eref{BFactn} to the Holst action \cite{holst, Parviol} for general relativity. 

Three parts comprise the rest of the section. In subsection \ref{subsecI1} we will  derive a semi-continuous BF action adapted to the simplicial discretization of the underlying manifold. This derivation has already been published previously \cite{hamspinfoam} (in a slightly more complicated form), it reappears here just to make the paper logically self-contained. The second part, subsection \ref{subsecI2} deals with the spinorial formalism. Once again, this just reviews results that have already been published. The last part presents our proposal: We introduce the discretized simplicity constraints \eref{simplcons}, and add them to the action. We will study these conditions in the form of the \emph{linear simplicity constraints} \cite{flppdspinfoam, LQGvertexfinite}, which replace \eref{simplcons} by an equivalent system of equations linear in the Pleba\'nski two-form $\Sigma_{\alpha\beta}$. This simplification is possible, however, only if we introduce an additional element to the theory: The time normals of the tetrahedra in the simplicial complex. We will argue that a consistent discretized theory must treat these normals as dynamical variables in the action. This idea will lead us to a proposal for a new action for discretized gravity in terms of first-order variables.
%-------------------------------------------------------------------------------%
\subsection{Semi-continuous BF action on a two-complex}\label{subsecI1}
\begin{figure}[h]
     %\centering\psfrag{a}{hallo}
     \includegraphics[width= 0.7\textwidth]{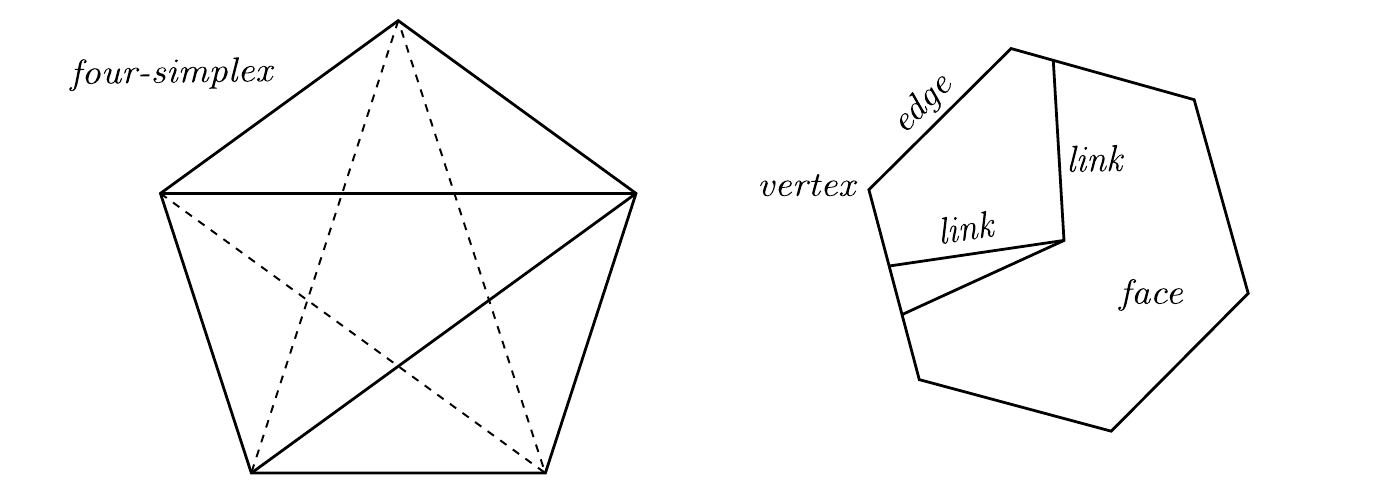}
     \caption{Left: %A simplicial discretization consists of several four-simplices glued among bounding tetrahedra. 
A four-simplex contains ten triangles $\tau,\tau',\dots$ and five tetrahedra $T, T',\dots$. On the right, there are the elements of the dual complex: A face $f$ is the two-dimensional surface dual to a triangle. Its boundary consists of edges $e, e',\dots$ connecting adjacent vertices $v_{ee'}$. Every vertex $v$ represents a four-simplex, and every edge is dual to a tetrahedron. Cutting a face into two halves creates two links: A link is a line connecting the center of the face with its boundary.}
     \label{fig1}
\end{figure}

\noindent
%-------------------------------------------------------------------------------%
This section deals with the discretization of the topological BF action \eref{BFactn} on a simplical decomposition of the spacetime manifold $M$. A simplicial complex (see figure \ref{fig1} for an illustration) consists of many four-simplices glued among their bounding tetrahedra. Each four-simplex contains five tetrahedra $T, T',\dots$ and ten triangles $\tau,\tau',\dots$. We then also need some elements of the dual complex: A four-simplex is dual to a vertex $v$, and each triangle is dual to a two-dimensional face $f$. The boundary of a face $f$ consists of several edges $e$, each of which represents an adjacent tetrahedron. 
Every triangle appears in many vertices, but a tetrahedron can only be in two of them. 

\begin{figure}[h]
     %\centering\psfrag{a}{hallo}
     \includegraphics[width= 0.34\textwidth]{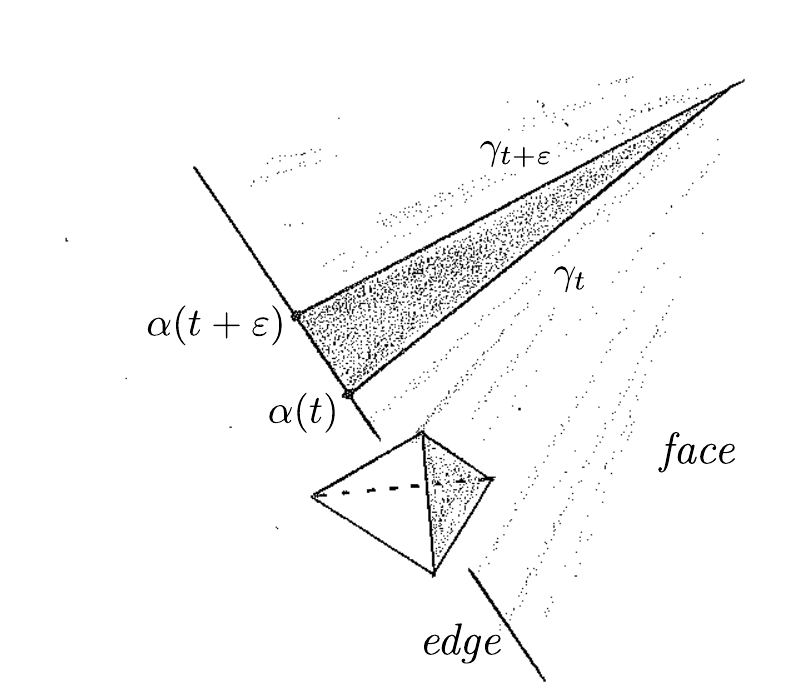}
     \caption{The boundary of a face $f$ is a loop $\alpha:[0,1]\rightarrow\partial f$, and an edge is a segment of this boundary between two vertices. Every face $f$ is dual to a triangle $\tau_f$, and every edge $e$ represents a tetrahedron $T_e$ bounded by four triangles---hence there are four faces adjacent to an edge. A link is a path $\gamma_t:[0,1]\rightarrow f, z\mapsto \gamma_t(z)$ connecting the point $\gamma_t(0)=\alpha(t)$ with the center $\gamma_t(1)=c=\tau_f\cap f$ of the face. The family of paths $\{\gamma_t\}_{t\in[0,1]}$ sweeps out the entire face, and only meets in $c$: $\forall t\neq t':\gamma_t\cap\gamma_{t'}=c$.} \label{fig2}
\end{figure}

The spacetime manifold $M$ carries an orientation, and so do all elements of the simplicial decomposition. The orientation of all faces $f$ and edges $e$ is arbitrary, the only constraint is that the relative orientation between a face $f$ and its dual triangle $\tau$ be one: If the pair $(X,Y)\in T_c\tau\subset T_cM$ has positive orientation in $\tau$, and  $(T,Z)\in T_c f$ is positively oriented in $f$, while $c\in f\cap \tau$ denotes the point common to both, then the quadruple $(T,X,Y,Z)$ of tangent vectors should have positive orientation in $T_cM$.
Given a simplicial discretization of $M$ we can then discretize the topological action \eref{BFactn} as a sum over faces (c.f. Thiemann's monograph \cite{thiemann} for a detailed explanation of the following formula):
\begin{equation}
S_{\text{BF}}[\Sigma,A]\approx\int_{M}\Pi_{\alpha\beta}\wedge F^{\alpha\beta}=\sum_{f:\text{faces}}\int_{\tau_f}\Pi_{\alpha\beta}\int_fF^{\alpha\beta}\equiv \sum_{f:\text{faces}}S_f,\label{BFactn1}
\end{equation}
where $\tau_f$ denotes the triangle dual to $f$. Equation \eref{BFactn1} breaks the local $SO(1,3)$ gauge symmetry. To restore gauge invariance, we introduce a two-parameter family of paths $\gamma_{p\rightarrow q}$ connecting any point $p\in \tau_f$ in the triangle with a point $q\in f$ in the face, compute the corresponding parallel propagator $h_\gamma=\mathrm{Pexp}(-\int_{\gamma}A)\in SO(1,3)$ and map all indices into a common frame. The underlying path $\gamma_{p\rightarrow q}$ splits into two halves. The first part lies inside the triangle $\tau_f$ and connects any $p\in\tau_f$ with the intersection $c=\tau_f\cap f$. The second part stays inside $f$ and goes out radially from the center of the face $c=\tau_f\cap f$ until it reaches $q\in f$. The result of this construction is a point-splitting regularization of the face contribution to the total action \eref{BFactn1}:
\begin{equation}
S_f={\int_f}\!\di t\,\di z{\int_{\tau_f}}\!\di x\,\di y\, \big[F_{q(t,z)}(\partial_t,\partial_z)\big]_{\alpha\beta}h\big[{(x,y)\rightarrow (t,z)}\big]{}^{\alpha}_{\phantom{\alpha}\mu}h\big[{{(x,y)\rightarrow (t,z)}}\big]{}^\beta_{\phantom{\beta}\nu}\big[\Pi_{p(x,y)}(\partial_x,\partial_y)\big]^{\mu\nu}
,\label{BFactn2}
\end{equation}
where $(x,y)\in[0,1)\times[0,1)$ and $(t,z)\in[0,1)\times[0,1)$ are positively oriented coordinates in $\tau_f$ and $f$ respectively, while, on the other hand, $h[(x,y)\rightarrow (t,z)]$ is just a shorthand notation for the holonomy along the connecting link $\gamma_{p(x,y)\rightarrow q(t,z)}$. Let us now use coordinates better adapted to the geometry of the problem. Any face $f$ has the topology of a disk, and we can thus turn the $(t,z)$ variables into \qq{polar} coordinates, such that for all $z\in[0,1]$ the set of points $\{q(t,z)\in f|t\in[0,1]\}$ defines a loop winding once around the face. For $z=0$ we shall sit at the boundary of the face $f$, while for $z\rightarrow 1$ all loops shall shrink to a point. We can then always identify this point with the intersection $c=\tau_f\cap f$ of the face with its dual triangle. In other words $\alpha:[0,1]\rightarrow M,t\mapsto \alpha(t)=q(t,z=0)$ is the loop bounding $f$, while $\gamma_t:[0,1]\rightarrow f\subset M,z\mapsto \gamma_t(z)=q(t,z)$ is a one-parameter family of links connecting the boundary of the face with its central point. See figure \ref{fig2} for an illustration.

We will now evaluate the integral \eref{BFactn2} in two steps. First of all, we look at the contribution from the $SO(1,3)$ curvature integrated over $f$. One of the most crucial steps of the paper is to write this integral as a one-dimensional line integral over the boundary $\partial f$ of $f$. This is possible thanks to the non-Abelian Stoke's theorem, which we can prove as follows. Consider first the link holonomy:
\begin{equation}
h_{\gamma_t}=\mathrm{Pexp}\big(-\int_{\gamma_t} A\big)\in SO(1,3),\label{linkhol}
\end{equation}
connecting the local $SO(1,3)$ frame at $\alpha(t)=q(t,z=0)\in\partial f$ in the boundary of $f$ with the frame at the center $c=\tau_f\cap f$ of the face. The $t$-derivative of this parallel propagator probes the curvature in the face $f$, and will immediately lead us to the non-Abelian Stoke's theorem. We compute this $t$-derivative by first going back to the defining differential equation for the link holonomy:
\begin{equation}
\frac{\di}{\di z}h_{\gamma_t(z)}=-A_{\gamma_t(z)}(\partial_z)h_{\gamma_t(z)},\label{holdef}
\end{equation}
with the initial condition $h_{\gamma_t(z=0)}=\mathds{1}$. We can now take the $t$-derivative of this equation, multiply the resulting expression by $h_{\gamma_t(z)}^{-1}$ from the left and integrate everything from $z=0$ to $z=1$. A partial integration eventually leaves us with the following expression:
\begin{equation}
h_{\gamma_t(1)}^{-1}\frac{\di}{\di t}h_{\gamma_t(1)}=-h_{\gamma_t(1)}^{-1}A_{\gamma_t(1)}(\partial_t)h_{\gamma_t(1)}+A_{\gamma_t(0)}(\partial_t)+\int_0^1\di z\,h_{\gamma_t(z)}^{-1}F_{\gamma_t(z)}(\partial_z,\partial_t)h_{\gamma_t(z)}.\label{smearedF0}
\end{equation}
This equation gives the \qq{velocity} of the holonomy under an infinitesimal variation $\gamma_t\rightarrow\gamma_t+\varepsilon\partial_z$ of the underlying link. At $z=1$ the vector field $\partial_t\in Tf$ vanishes, because all links come together at the center of the face: $ \forall t:\gamma_t(0)=c=\tau_f\cap f$. The first term of equation \eref{smearedF0} therefore vanishes, which eventually leaves us with the following equation:
\begin{equation}
F_f(t):=\int_0^1\di z\,h_{\gamma_t(z)}^{-1}F_{\gamma_t(z)}(\partial_z,\partial_t)h_{\gamma_t(z)}=h_{\gamma_t(1)}^{-1}\frac{D}{\di t}h_{\gamma_t(1)}=h_{\gamma_t(1)}^{-1}\Big(\frac{\di}{\di t}h_{\gamma_t(1)}-h_{\gamma_t(1)}A_{\gamma_t(0)}(\partial_t)\Big),\label{smearedF}
\end{equation}
where $D/{\di t}$ is the covariant $SO(1,3)$ $t$-derivative.  Notice also, that  the $\mathfrak{so}(1,3)$ element $F_f(t)$ transforms homogeneously under local gauge transformations in the fibre over $\alpha(t)$ (with $\alpha(t)=\gamma_t(0)$ parametrizing the loop bounding the face). To be more precise, if $\tilde{A}=\Lambda^{-1}\di \Lambda + \Lambda^{-1}A\Lambda$ is a gauge equivalent connection, with $\Lambda:M\rightarrow SO(1,3)$ denoting the gauge element, then $F_f(t)$ transforms according to: 
\begin{equation}
\tilde{F}_f(t)=\Lambda^{-1}\big(\alpha(t)\big)F_f(t)\Lambda\big(\alpha(t)\big)\in\mathfrak{so}(1,3).
\end{equation}\label{Ftrans}
Equation \eref{smearedF} generalizes the usual Stoke's theorem for Abelian one-forms. Let us explain this more carefully: We map $F_f(t)$ into the frame at the center of the face\footnote{In the following, all variables carrying an \qq{undertilde} will always belong to the fibre over the center of the corresponding face.} $\utilde{F}_f(t):=h_{\gamma_t}F_f(t)h_{\gamma_t}^{-1}\in \mathfrak{so}(1,3)$, and integrate the resulting expression over all values of $t$:
\begin{equation}
\int_f\di t\,\di z\,h_{\gamma_t(1)}h_{\gamma_t(z)}^{-1}F_{\gamma_t(z)}(\partial_z,\partial_t)h_{\gamma_t(z)}h_{\gamma_t(1)}^{-1}=\int_{\partial f}\di t\,\frac{D}{\di t}h_{\gamma_t(1)}h_{\gamma_t(1)}^{-1}\in\mathfrak{so}(1,3).\label{Stokesthrm}
\end{equation}
If we now replace, for just a second, the gauge group $SO(1,3)$ by the Abelian group $U(1)$, we immediately see that equation \eref{Stokesthrm} turns into the usual Stoke's theorem for a $U(1)$ gauge potential $A$: $\int_{f}F=\int_f \di A=\int_{\partial f}A$, hence justifying the terminology of a non-Abelian Stoke's theorem.

The smeared curvature tensor $F_f(t)$ as defined in \eref{smearedF} gives us the first contribution to the regularized integral \eref{BFactn2}. The other contribution comes from the integral of the momentum variable $\Pi^{\alpha\beta}$ over the dual triangle. Once again this integral requires a family of holonomies mapping the $(\alpha, \beta)$ indices into a common frame, for we cannot add internal vectors sitting at separate points in a gauge invariant manner. The same is also true for the contraction between the smeared momentum variable $\propto\int_{\tau}\Pi^{\alpha\beta}$ and the curvature integral $F_f(t)$ over a link. The contraction can only happen in a common frame. The Lie algebra element $\ou{[F_f(t)]}{\beta}{\alpha}$ belongs to the $\mathfrak{so}(1,3)$-fibre over $\alpha(t)\in\partial f$, so we should also map $\ou{\Pi}{\alpha}{\beta}$ into that frame. We thus introduce a three-parameter family of paths $\gamma(t,x,y)$ linking any point $p(x,y)\in\tau_f$ in the triangle with the point $\alpha(t)$ in the boundary $\partial f$ of the face. Once again, $\gamma(t,x,y)$ consists of two parts. The first half lies inside the triangle and connects the point $p(x,y)\in\tau_f$ with the intersection $c=f\cap \tau_f$, the other half follows $\gamma_t^{-1}$ connecting $c$ with $\alpha(t)$ in the boundary of the face $f$. We can thus introduce the smeared momentum in the frame over $\alpha(t)$:
\begin{equation}
\Pi^{\alpha\beta}_f(t)=\int_{\tau_f}\di x\,\di y\,\ou{[h_{\gamma(t,x,y)}]}{\alpha}{\mu}\ou{[h_{\gamma(t,x,y)}]}{\beta}{\nu}\big[\Pi_{p(x,y)}(\partial_x,\partial_y)\big]^{\mu\nu}.\label{fluxdef}
\end{equation}

Combining \eref{fluxdef} with the curvature in a link \eref{smearedF} we thus get the following integral for the face contribution \eref{BFactn2} to the discretized topological action \eref{BFactn1}:
\begin{equation}
S_f=-\int_{\partial f}\di t\,\Big[h_{\gamma_t(1)}^{-1}\frac{D}{\di t}h_{\gamma_t(1)}\Big]_{\alpha\beta}\Pi^{\alpha\beta}_f(t),\label{BFactn3}
\end{equation}
with $D/\di t$ denoting the covariant derivative along the boundary of the face $f$. Equation \eref{BFactn3} gives the contribution to the action from a single face $f$, and we obtain the discretized BF action for the entire four-dimensional manifold simply by summing over all faces in the simplicial complex, i.e. $S_{\mathrm{BF}}=\sum_{f:\text{faces}}S_f$ according to \eref{BFactn1}.  The Lagrangian of the action \eref{BFactn3} is nothing but a covariant symplectic potential for the phase space $T^\ast SO(1,3)$. The absence of a non-trivial Hamiltonian reflects the topological nature of BF theory. Only additional constraints can change this situation. (We do not expect the appearance of a \qq{true Hamiltonian}, because that would always violate the residual one-dimensional coordinate invariance $t\rightarrow \tilde{t}(t)$ of the action.)

%-------------------------------------------------------------------------------%
\subsection{Spinors}\label{subsecI2}\noindent
%-------------------------------------------------------------------------------%
The discretized action \eref{BFactn3} is a functional of three elements: The gravitational flux $\Pi^{\alpha\beta}_f$ as defined in \eref{fluxdef}, the link holonomy $h_{\gamma}$ \eref{linkhol}, and the $SO(1,3)$ connection hiding in the covariant derivative ${D}/{\di t}$. The decomposition of the covariant derivative $\tfrac{D}{\di t}V^\alpha=\partial_tV^\alpha+[A_{\alpha(t)}(\partial_t)]{}^\alpha{}_\beta V^\beta$ into the partial $t$-derivative and the $SO(1,3)$ connection $\ou{A}{\alpha}{\beta}$ immediately leads us to the symplectic structure of the theory. The corresponding phase space is the cotangent bundle of the Lorentz group, and we can thus attach to each link in the discretization the phase space $T^\ast SO(1,3)$ of holonomies \eref{holdef} and fluxes \eref{fluxdef}. 

The phase space $T^\ast SO(1,3)$ has no canonical coordinates globally, which makes calculations unnecessarily complicated. We can, however, embed $T^\ast SO(1,3)$ into a linear vector space with global Darboux coordinates. The construction starts from the decomposition of the gravitational flux \eref{fluxdef} into its selfdual ($\Pi^{AB}_f$) and anti-selfdual ($\bar\Pi^{\bar A\bar B}_f$) components:
\begin{equation}
\Pi_f^{\alpha\beta}(t)\equiv\Pi^{AA'BB'}_f(t)=-\bar\epsilon^{A'B'}\Pi_f^{AB}(t)+\CC,\label{fluxspin}
\end{equation} 
where we have implicitly used the Infeld--van der Waerden symbols\footnote{If we want to use an explicit matrix representation we can identify $\sigma_{0}$ with the identity, and the spatial components $\sigma_{i}$ ($i=1,2,3$) with the Pauli spin matrices.} $\ou{\sigma}{AA'}{\alpha}$ to identify any Lorentz vector $X^\alpha$ with an anti-Hermitian\footnote{The anti-Hermiticity of $X^{AA'}$ is a result of our choice $(-+++)$ for the metric signature.} $2\times 2 $ matrix: $X^{AA'}=\frac{\I}{\sqrt{2}}\ou{\sigma}{AA'}{\alpha}X^\alpha$. Elements $v^A\in\C^2$ carry abstract spinor indices $A,B,C,\dots$ transforming under the fundamental spin $(\tfrac{1}{2},0)$ representation of $SL(2,\C)$. The primed indices $A',B',C',\dots$ belong to the complex conjugate vector space $\bar\C^2$, and carry a representation of the irreducible spin  $(0,\tfrac{1}{2})$ representation. There is also the algebraic dual $[\C^2]^\ast\ni v_A$, and the anti-symmetric $\epsilon$-tensor $\epsilon_{AB}=-\epsilon_{BA}$ moves the corresponding indices:
$ v_A=\epsilon_{BA}v^B$ and $v^A=\epsilon^{AB}v_B$,
with $\epsilon^{AB}$ denoting the inverse of $\epsilon_{AB}$: $\epsilon^{AC}\epsilon_{BC}=\delta^A_B$. Finally, $\CC$ denotes the complex conjugate of everything preceding.

The flux $\Pi_f^{\alpha\beta}$ is anti-symmetric, hence $\Pi^{AB}_f=\Pi^{BA}_f$ is symmetric, and has therefore two complex eigenspinors, which we call $\omega^A_f$ and $\pi_f^A$:
\begin{equation}
\Pi^{AB}_f(t)=-\frac{1}{2}\omega^{(A}_f(t)\pi^{B)}_f(t),\label{fluxpar}
\end{equation}
where $(A\dots B)$ denotes symmetrization of all intermediate indices. The eigenspinors transform under local $SL(2,\C)$ gauge transformations at the boundary of $f$, and we thus think of them as local sections of a spinor bundle over the boundary $\alpha(t)\in\partial f$ of the underlying face. Furthermore, the spinors should be both periodic and differentiable in $t$, e.g.: $\omega^{A}(0)=\omega^{A}(1)$.

The link holonomy $\ou{[h_{\gamma_t}]}{\alpha}{\beta}=\ou{[h_{\gamma_t}]}{A}{B}\ou{[\bar{h}_{\gamma_t}]}{A'}{B'}$ \eref{linkhol} maps the $\C^2$-fibers over the boundary $\alpha(t)\in\partial f$ of $f$ into the common fibre over the center $c=\tau_f\cap f$ of the underlying face. We can thus map all spinors $\pi^A_f(t)$ and $\omega^A_f(t)$ into the same frame, thus obtaining yet another pair $(\utilde{\pi}^A,\utilde{\omega}^A)$ of spinors:
\begin{equation}
\utilde{\pi}^A_f(t):=\ou{[h_{\gamma_t}]}{A}{B}\pi^B_f(t),\quad
\utilde{\omega}^A_f(t):=\ou{[h_{\gamma_t}]}{A}{B}\omega^B_f(t).\label{tildedef}
\end{equation}
The spinors often come in pairs, it is therefore useful to introduce the twistors $Z$ and $\utilde{Z}$:
\begin{equation}
Z:=(\bar{\pi}_{A'},\omega^A),\quad\text{and:}\quad\utilde{Z}:=(\utilde{\bar{\pi}}_{A'},\utilde{\omega}^A).
\end{equation}

Consider now the $SL(2,\C)$-invariant contraction\footnote{The relative position of the index $f$ labeling the faces has no meaning, we put it wherever there is enough free space left: $\pi_A^f=\epsilon_{BA}\pi^B_f$.} of the spinors:
\begin{equation}
E_f=\pi_A^f\omega^A_f=\utilde{\pi}_A^f\utilde{\omega}^A_f.
\end{equation}
If $E_f=0$, the spinors are linearly dependent and the corresponding triangle $\tau_f$ is null: $\Pi^{\alpha\beta}_f\Pi_{\alpha\beta}^f=0=\ast\Pi_{\alpha\beta}\Pi^{\alpha\beta}$; $E_f\neq 0$, on the other hand, implies that the triangle is non-degenerate, and allows us to use the spinors $(\pi^A_f,\omega^A_f)$ as a complex basis in $\C^2$. Let us agree on this additional constraint, since we will soon restrict ourselves to spacelike triangles
 anyhow. We thus have a basis $(\pi^A_f,\omega^A_f)$ attached to the boundary of $f$, and a basis $(\utilde{\pi}^A_f,\utilde{\omega}^A_f)$ in the frame of the center of the face $f$. We can then invert equation \eref{tildedef} and write the $SL(2,\C)$-holonomy $\ou{[h_{\gamma_t}]}{A}{B}$ as a function of the spinors:
\begin{equation}
{\big[h_{\gamma_t}\big]}^{A}_{\phantom{A}B}=\frac{\utilde{\omega}^A(t)\pi_B(t)-\utilde{\pi}^A(t)\omega_B(t)}{\sqrt{E(t)}\sqrt{\utilde{E}(t)}},\label{holpar}
\end{equation}
where we have dropped the $f$-label to keep the equation simple, while $E=\pi_A\omega^A$ and $\utilde{E}=\utilde{\pi}_A\utilde{\omega}^A$. 

We now have all the component parts in hand to write the discretized BF action \eref{BFactn3} fully in terms of our spinorial variables. We go back to the face action \eref{BFactn3}, and insert the spinorial parametrization (\ref{fluxpar}, \ref{fluxspin}, \ref{holpar}) of both flux  $\Pi^{\alpha\beta}=-{\bar\epsilon}^{A'B'}\Pi^{AB}+\CC$ and holonomy $\ou{h}{\alpha}{\beta}=\ou{h}{AA'}{BB'}=\ou{h}{A}{A'}\ou{\bar{h}}{A'}{B'}$. This brings the action into the following form:
\begin{equation}
S_{f}=\frac{1}{2}\int_{0}^{1}\di t\Big[\omega_{A}\frac{D}{\di t}\pi^{A}+\pi_{A}\frac{D}{\di t}\omega^{A}-
	\frac{E}{\utilde{E}}\big(\utilde{\pi}_{A}\utilde{\dot{\omega}}^{A}-\utilde{\dot{\pi}}_{A}\utilde{\omega}^{A}\big)\Big]+\CC,\label{spinactn}
\end{equation}
where the covariant $SO(1,3)$ derivative has turned into the corresponding $SL(2,\C)$ derivative: $\frac{D}{\di t}\pi^A=\frac{\di}{\di t}\pi^A+[A_{\alpha(t)}(\partial_t)]^A{}_B\pi^B$, with $\ou{A}{A}{B}$ denoting the selfdual component of the $SO(1,3)$ connection $\ou{A}{\alpha}{\beta}$. The covariant derivative of the $(\utilde{\omega}^{A},\utilde{\pi}^{B})$-spinors, on the other hand, is just an ordinary derivative, e.g.: $\frac{D}{\di t}\utilde{\pi}^A=\utilde{\dot{\pi}}^A$. The reason is the same that brought us from equation \eref{smearedF0} to equation \eref{smearedF}: The $(\utilde{\omega}^{A},\utilde{\pi}^{B})$-spinors belong to the frame at the center of the face $f$. At the central point $c=\tau_{f}\cap f$ all the links come together radially $\forall t$: $c=\gamma_{t}(1)$. The vector field $\partial_{t}$ thus vanishes at $c=\gamma_{t}(1)$, hence $\frac{D}{\di t}\utilde{\pi}^A(t)=\utilde{\dot{\pi}}^A(t)+\ou{[A_{c}(\partial_{t})]}{A}{B}\utilde{\pi}^{B}(t)=\utilde{\dot{\pi}}^A(t)$.

The $({\pi},\omega)$- and $(\utilde{\pi},\utilde{\omega})$-spinors are not independent, for there is the $SL(2,\C)$ holonomy that maps one into the other (see equation \eref{tildedef}). If we want to use the spinors as our fundamental configuration variables, we need a constraint that guarantees the existence of a linking $SL(2,\C)$ holonomy. The \emph{area-matching constraint} 
 \begin{equation}
\varDelta:=\utilde{\pi}_{A}\utilde{\omega}^A-{\pi}_{A}{\omega}^A=\utilde{E}-E\label{areamatch}
\end{equation}
does the job: Equation \eref{areamatch} requires that the spinors have equal $SL(2,\C)$ norm $\pi_A\omega^A\in\C$, and reduces therefore the matrix \eref{holpar} down to an $SL(2,\C)$ element. Hence, we introduce a Lagrange multiplier $\zeta:\partial f\rightarrow \C$ and add the area-matching constraint \eref{areamatch} to the action. Performing a partial integration eventually leaves us with the following action on each face:
\begin{equation}
S_{f}[Z,\utilde{Z},A,\zeta]=\int_{0}^{1}\di t\Big[\pi_{A}\frac{D}{\di t}\omega^{A}-\utilde{\pi}_{A}\frac{\di}{\di t}\utilde{\omega}^{A}
+\zeta\big(\pi_{A}\omega^{A}-\utilde{\pi}_{A}\utilde{\omega}^{A}\big)
\Big]+\CC\label{spinBF},
\end{equation}
where there are no boundary terms appearing, because the spinors are periodic in $t$, e.g.: $\omega^{A}(0)=\omega^{A}(1)$. The action for the entire discretized manifold is the sum over the contributions \eref{spinBF} from each individual face, a sum which yields the following twistorial BF action:
\begin{align}\nonumber
S_{\mathrm{BF}}[Z_{f_1},Z_{f_2},&\dots;\utilde{Z}_{f_1},\utilde{Z}_{f_2},\dots;\zeta_{f_1},\zeta_{f_2},\dots;\Lambda_{e_1},\Lambda_{e_2},\dots]=\sum_{f:\text{faces}}S_f=\\
&=\sum_{f:\text{faces}}\oint_{\partial f}\Big[\pi_A^f D\omega^A_f-\utilde{\pi}_A^f\di\utilde{\omega}^A_f+\zeta_f\big(\pi_{A}^f\omega^{A}_f-\utilde{\pi}^f_{A}\utilde{\omega}^{A}_f\big)\Big]+\CC\label{spinBF2}
\end{align}
This action is a functional of two twistor fields $(Z_f,\utilde{Z}_f)$  and a complex-valued Lagrange multiplier\footnote{We have absorbed here the integration measure $\di t$ into the definition of $\zeta_f$; which is therefore a one-form on the boundary of $f$.} $\zeta_f$ for each face, and an $SL(2,\C)$ connection $\Lambda_e$ on the edges:
\begin{equation}
D\pi^{A}_{f}\Big|_{e(t)}=\di t\Big[\frac{\di}{\di t}{\pi}^A_{f}+\ou{[\Lambda_{e}(t)]}{A}{B}\pi^{B}_{f}\Big]_{e(t)},
\end{equation}
where $\Lambda_e(t)\in\mathfrak{sl}(2,\C)$ denotes the selfdual connection contracted with the tangent vector $\dot{e}\in T_{e(t)}\partial f\subset T_{e(t)}M$ of the underlying edge $e:[0,1]\rightarrow\partial f, t\mapsto e(t)$:
\begin{equation}
\ou{\Lambda_e(t)}{A}{B}:=\ou{[A_{e(t)}(\dot{e})]}{A}{B}\in \mathfrak{sl}(2,\C).
\end{equation}

Before we go on, and add the simplicity constraints to the action \eref{BFactn2}, let us briefly look at the equations of motion for the topological theory.  First of all, we compute the variation of the spinors in a face, and immediately find them to diagonalize the parallel transport around that face: We get for e.g. $\pi^A_f$ that $\frac{D}{\di t}\pi^A_t=\zeta_f\pi^A_f$. We can now trivially integrate these evolution equations to find:
\begin{equation}
\begin{split}
\pi_f^A(t)=\E^{+\int_{\alpha_t}\!\zeta_f}\ou{[h_{\alpha_t}]}{A}{B}\pi_f^B(0),\qquad&\utilde{\pi}_f^A(t)=\E^{+{\int_{\alpha_t}}\!\zeta_f}\utilde{\pi}_f^A(0),\\
\omega_f^A(t)=\E^{-\int_{\alpha_t}\!\zeta_f}\ou{[h_{\alpha_t}]}{A}{B}\omega_f^B(0),\qquad&\utilde{\omega}_f^A(t)=\E^{-{\int_{\alpha_t}}\!\zeta_f}\utilde{\omega}_f^A(0),\label{spinevolv}
\end{split}
\end{equation}
where $\alpha:[0,1]\rightarrow\partial f$ parametrizes the boundary of $f$, hence $\alpha(0)=\alpha(1)$, while $\alpha_t$ is the segment $\alpha_t=\{\alpha(s)|0\leq s\leq t\}$  with $h_{\alpha_t}=\mathrm{Pexp}(-\int_{\alpha_t}A)\in SL(2,\C)$ denoting the corresponding holonomy from $\alpha(0)$ to $\alpha(t)$. 
The periodic boundary conditions for the spinors, e.g. $\pi^A_f(0)=\pi^A_f(1)$, lead us to the flatness of the connection:
\begin{subequations}\begin{align}
\exists n_f\in \Z:\oint_{\partial f}\zeta_f=2\pi \I n_f,\label{BFflata}\\
\ou{[h_{\partial f}]}{A}{B}=\mathrm{Pexp}\Big(-\oint_{\partial f}A\Big)^{A}_{\phantom{A}B}=\delta^A_B.\label{BFflatb}
\end{align}\label{BFflat}
\end{subequations}
The parallel transport around the loop bounding a face is therefore just the identity, which should not come as a surprise:  We have started from the topological BF theory \eref{BFactn} in the continuum; its action variation immediately yields the vanishing of curvature, hence the connection is flat, implying that any spinor is transported into itself once we go around a loop.

We are now left to study the variation of the $SL(2,\C)$ connection $\Lambda_e(t)=A_{e(t)}(\partial_t)\in\mathfrak{sl}(2,\C)$ along the edges. Imagine first a single edge $e$ in the simplicial discretization. This edge will bound four adjacent faces, because any edge is dual to a tetrahedron containing \emph{four} triangles, each one of which is dual to an adjacent face (see figure \ref{fig2} for an illustration). The variation $\delta\Lambda_e(t)$ of the connection must therefore appear in four terms in the action---one for each adjacent face. Since $\delta\frac{D}{\di t}\pi^A_f(t)=\ou{[\delta\Lambda_e(t)]}{A}{B}\pi^B_f(t)$ these terms immediately combine to form the Gauß constraint:
\begin{equation}
\forall e(t):\quad\smashoperator{\sum_{f:\partial f\ni e(t)}}\varepsilon(e,f)\,\pi_{(A}^f\omega_{B)}^f\Big|_{e(t)}=-2\smashoperator{\sum_{f:\partial f\ni e(t)}}\varepsilon(e,f)\Pi^f_{AB}\Big|_{e(t)}=0,
\label{BFGauss}\end{equation}
where $\varepsilon(e,f)$ denotes the relative orientation\footnote{The induced orientation of $\partial f$ may not match the orientation of $e$, in which case $\varepsilon(e,f)=-1$, whilst otherwise $\varepsilon(e,f)=1$.} between $f$ and $e$, while $(A\dots B)$ symmetries all intermediate indices. Once again equation \eref{BFGauss} just proves the consistency of the discretization: Our starting point was the topological BF action \eref{BFactn1} in the continuum, its connection variation yields the constraint: $D\Pi^{\alpha\beta}=0$, hence also $D\Pi^{AB}=0$. We can then integrate this three-form over a tetrahedron $T$. Employing the non-Abelian Stoke's theorem eventually yields a surface integral over the four bounding triangles, schematically: {\small $\int_TD\Pi^{\alpha\beta}=\int_{\partial T}\Pi^{\alpha\beta}=\sum_{\tau:\tau\subset\partial T}\int_{\tau}\Pi^{\alpha\beta}$}, where all internal indices are to be mapped into a common frame. Each of these integrals defines a momentum flux {\small$\Pi^{\alpha\beta}_f$} just as in \eref{fluxdef}. In other words: It does not matter if we first start from the continuum theory, compute the variation of the action, and then discretize, or first discretize the action and then derive the equations of motion from the variation of the discretized action. In both cases we end up with the same set of equations, i.e. the flatness constraints \eref{BFflat} and the Gauß law \eref{BFGauss}.
%-------------------------------------------------------------------------------%
\subsection{Discretised simplicity constraints}\label{subsecI3}\noindent
%-------------------------------------------------------------------------------%
  The last section studied the BF action discretized over a simplicial decomposition of the four-dimensional spacetime manifold. The resulting discretized action \eref{spinBF2} is a sum over one-dimensional integrals---one for each edge in the discretization. The fundamental configuration variables are the $(\utilde{\pi},\utilde{\omega})$- and $(\pi,\omega)$-spinors parametrizing the holonomy-flux variables (\ref{fluxpar}, \ref{holpar})

We are, however, not just interested in the topological BF theory by itself. The goal is to follow the Pleba\'{n}ski principle and develop a spinorial version of first-order Regge calculus. What we thus need are the simplicity constraints \eref{simplcons} that guarantee the existence of a tetrad $e_\alpha:\Sigma_{\alpha\beta}=e_\alpha\wedge e_\beta$ and reduce, therefore, the topological BF action \eref{BFactn} to the Holst action for general relativity. Having a discretized BF action at hand, we are now ready for the next step: To discretize the simplicity constraints and add them to the action. %But we are working with a simplicial action \eref{spinBF2}, and thus first need a suitable discretization of the simplicity constraints before we can add them to the action.

The simplicity constraints impose the geometricity of the Pleba\'nski two-form $\Sigma_{\alpha\beta}$. If the geometry is locally flat, these constraints further imply that any tetrahedron $T$ (dual to an edge $e$) has a unique normal $n^\alpha_e\in\R^4$:
\begin{equation}
\Sigma_{\alpha\beta}^fn^\beta_e\equiv\Sigma_{AA'BB'}^fn^{BB'}_e-\CC=-\Sigma^f_{AB}n_{e\,A'}^B-\CC=0,\label{linsimpl}
\end{equation}
where $\{\Sigma_{\alpha\beta}^f\}_{f:\partial f\supset e}$ are the $\mathfrak{so}(1,3)$-fluxes through the tetrahedron's four bounding triangles, and  $\Sigma_{AB}$ is the selfdual part of $\Sigma_{\alpha\beta}\equiv\Sigma_{AA'BB'}=-\epsilon_{AB}\bar\Sigma_{A'B'}+\CC$ The reverse is also true: If equation \eref{linsimpl} and the Gauß law $\sum_{f:\partial f\supset e}\Sigma^f_{\alpha\beta}=0$ hold true for all edges $e$ and adjacent faces $f$, then this suffices\footnote{The proofs can be found in \cite{LQGvertexfinite, flppdspinfoam}.} to reconstruct a locally flat metric around every four-simplex in the discretization. For any locally flat geometry, the \emph{linear simplicity constraints} \eref{linsimpl} together with the Gauß law are therefore equivalent to the continuous simplicity constraints \eref{simplcons} imposing the existence of a tetrad: $\exists e_\alpha:\Sigma_{\alpha\beta}-e_\alpha\wedge e_\beta=0$. The Gauß law appears already in the list of equations of motion derived from the BF action, and it should thus suffice to add just the linear simplicity constraints \eref{linsimpl} for all edges $e$ and adjacent faces $f$ to our discretized BF action \eref{spinBF2}. In the following we will restrict ourselves to spacelike tetrahedra, hence $n^\alpha n_\alpha=-1$, and without loss of generality $n^{\alpha}$ be always future pointing.
It should also be possible to generalize our formalism along the lines of \cite{Mingyitwist} and \cite{Conrady:2010kc,Conrady:2010vx,Conrady:2010sx} so as to allow for also null or timelike tetrahedra. 

To impose the linear simplicity constraints \eref{linsimpl} on the phase space of the theory, we first need the relation between the momentum variable $\Pi_{\alpha\beta}$ and the Pleba\'nski two-form $\Sigma_{\alpha\beta}$. Going back to \eref{BFactn} we find:
\begin{equation}
\Sigma_{\alpha\beta}=-\frac{2\ellp^2}{\hbar}\frac{\beta}{\beta^2+1}\Big(\beta\ast\!\Pi_{\alpha\beta}+\Pi_{\alpha\beta}\Big),\quad\text{equally:}\quad
\Sigma_{AB}=\frac{2\ellp^2}{\I\hbar}\frac{\beta}{\beta+\I}\Pi_{AB}.\label{pisigma}
\end{equation}
If the tetrahedra are all spacelike, the fluxes $\Pi^f_{\alpha\beta}$ through their bounding triangles are non-degenerate: $\Pi^f_{\alpha\beta}\Pi_f^{\alpha\beta}\neq 0$, thus $\Pi^f_{AB}\Pi^{AB}_f\neq 0$ and also $\pi_A^f\omega^A_f\neq 0$. The non-degeneracy of the spinors, i.e. $\pi_A\omega^A\neq 0$, implies that they form a basis in $\C^2$, which then also guarantees that the complex null-vectors $\{\ell^\alpha\equiv\pi^A\bar\pi^{A'}, k^\alpha\equiv\omega^A\bar\omega^{A'},m^\alpha\equiv\I\pi^A\bar\omega^{A'},\bar{m}^\alpha\equiv\I\omega^A\bar\pi^{A'}\}$ are linearly independent and span all of complexfied Minkowski space $\C^4$. We can now go back to the simplicity constraint $\Sigma_{\alpha\beta}^fn^\beta_e$ and contract the free Lorentz index with these basis vectors, thus obtaining a set of equivalent constraints:
\begin{subequations}
\begin{align}
&V_f=\frac{\I}{\beta+\I}\pi_A^f\omega^A_f+\CC\stackrel{!}{=}0,\label{spinsimplb}\\
&W_{ef}=n^{AA'}_e\pi_A^f\bar\omega_{A'}^f\stackrel{!}{=}0,\label{spinsimpla}
\end{align}\label{spinsimpl}
\end{subequations}
where we have employed the spinorial parametrization of the selfdual flux $2\Pi_{AB}=-\omega_{(A}\pi_{B)}$, i.e. equation \eref{fluxpar}. Notice that $W_{ef}$ is complex, while $V_f$ is real, so we are dealing with three constraints for each face $f$.

\subparagraph*{} At this point it is now useful to study the little group $SU(2)_n=\{\ou{h}{A}{B}\in SL(2,\C):n^{AA'}=\ou{h}{A}{B}\ou{\bar{h}}{A'}{B'}n^{BB'}\}$ preserving the time normal $n^\alpha$, its Lie algebra $\mathfrak{su}(2)_n$, and to introduce some additional geometrical structures. Given a future-oriented time-normal $n^\alpha$, $n^\alpha n_\alpha=-1$ we can define a three-metric $h_{\alpha\beta}$ together with the corresponding volume element $\epsilon_{\alpha\beta\mu}$:
\begin{equation}
h_{\alpha\beta}=\eta_{\alpha\beta}+n_\alpha n_\beta,\qquad \epsilon_{\alpha\beta\mu}=\epsilon_{\nu\alpha\beta\mu}n^\nu.
\end{equation}
On the level of the corresponding Lie algebra, i.e. $\mathfrak{sl}(2,\C)$ respectively $\su(2)_n$, there is a similar decomposition: Starting from the Infeld--van der Waerden symbols $\ou{\sigma}{AA'}{\alpha}$, we can define the generators $\ou{\tau}{A}{B\mu}$ of the $SU(2)_n$ little group:
\begin{equation}
\ou{\tau}{A}{B\mu}=\frac{1}{4\I}\Big(\ou{\sigma}{AC'}{\mu}\bar{\sigma}_{C'B\nu}-\ou{\sigma}{AC'}{\nu}\bar{\sigma}_{C'B\mu}\Big)n^\nu.\label{Pmatrx}
\end{equation}
These matrices generalize the Pauli spin matrices: They are anti-Hermitian\footnote{That is: $\delta_{AA'}\ou{\bar\tau}{A'}{B'\alpha}\delta^{BB'}=-\ou{\tau}{B}{A\alpha}.$} with respect to the $SU(2)_n$ metric:
\begin{equation}
\delta_{AA'}=\sigma_{AA'\alpha}n^\alpha,
\end{equation}
and obey the generalized Pauli identity:
\begin{equation}\label{Paulident}
\ou{[\tau_\mu\tau_\nu]}{A}{B}\equiv\ou{\tau}{A}{C\mu}\ou{\tau}{C}{B\nu}=
-\frac{1}{4}\delta^A_Bh_{\mu\nu}+\frac{1}{2}\ou{\epsilon}{\alpha}{\mu\nu}\ou{\tau}{A}{B\alpha}.
\end{equation}
Moreover, the matrices $\tau_\mu$ span both $\mathfrak{su}(2)_n$ and $\mathfrak{sl}(2,\C)$: Given any Lie algebra element $\ou{\phi}{A}{B}\in\mathfrak{su}(2)_n$ we can define its components $\phi^\alpha\in\R^4$ through the decomposition $\ou{\phi}{A}{B}=\phi^\alpha\ou{\tau}{A}{B\alpha}$, while for any $\ou{\phi}{A}{B}\in\mathfrak{sl}(2,\C)$ the components $\phi^\alpha$ in $\ou{\phi}{A}{B}=\phi^\alpha\ou{\tau}{A}{B\alpha}$ are generally complex, i.e. $\phi^\alpha\in\C^4$. The Pauli matrices are purely spatial ($\ou{\tau}{A}{B\alpha}n^\alpha=0$) hence $\phi^\alpha n_\alpha=0$ without loss of generality.

\subparagraph*{} Before we go on, let us close this section with one more remark. In the literature \cite{flppdspinfoam, LQGvertexfinite, zakolec} there often appears yet another equivalent form of the simplicity constraints \eref{linsimpl}. Given the tetrahedron's time normal $n^\alpha_e$ we can define the electric and magnetic parts of the momentum variable:
\begin{equation}
K_\alpha^{ef}=2\Pi_{\alpha\beta}^fn^\beta_e,\qquad
L_\alpha^{ef}=2\ast\!\Pi_{\alpha\beta}^fn^\beta_e,\label{LKdef}
\end{equation}
where $\ast\Pi_{\alpha\beta}=\frac{1}{2}\uo{\epsilon}{\alpha\beta}{\mu\nu}\Pi_{\mu\nu}$ denotes again the Hodge dual. Notice also that the fluxes $L_\alpha^{ef}$ and $K_\alpha^{ef}$ carry two labels, $f$ refers to the triangle under consideration: $K_\alpha^{ef}$ and $L_\alpha^{ef}$ are the electric and magnetic components of the two-form $\Pi_{\alpha\beta}$ smeared over the triangle $\tau_f$ dual to the face $f$. The edge-lables $e\subset \partial f$, on the other hand, come from the tetrahedron's (every tetrahedron is dual to an edge $e$) time-normal $n^\alpha_e\in\R^4$, with respect to which we have decomposed $\Pi_{\alpha\beta}^f$ into its magnetic and electric parts. Let us now come back to the linear simplicity constraints: $n^\alpha_e\Sigma_{\alpha\beta}^f=0$. Equation \eref{pisigma} tells us that the momentum $\Pi_{\alpha\beta}$ is linearly related to the two-form $\Sigma_{\alpha\beta}$, thus turning the linear simplicity constraints into a relation between the generators $L_\alpha$ and $K_\alpha$:
\begin{equation}
\Sigma_{\alpha\beta}^fn^\beta_e=-\frac{\ellp^2}{\hbar}\frac{\beta}{\beta^2+1}\Big(\beta L_\alpha^{ef}+K_\alpha^{ef}\Big)=0.\label{KLsimpl}
\end{equation}
which is, in fact, the most popular way to write the linear simplicity constraints \cite{flppdspinfoam, LQGvertexfinite, zakolec, Bianchientropy} in the literature.
%-------------------------------------------------------------------------------%
\subsection{Emergence of the Ashtekar--Barbero connection}\noindent
%-------------------------------------------------------------------------------%
The simplicity constraints \eref{linsimpl} attach a direction $n^\alpha\in\R^4$ to every edge. These directions, the time-normals of the elementary tetrahedra, break the local $SL(2,\C)$ gauge symmetry down to the little group $SU(2)_n=\{\ou{h}{A}{B}\in SL(2,\C):n^{AA'}=\ou{h}{A}{B}\ou{\bar{h}}{A'}{B'}n^{BB'}\}$.  This, in turn reduces the selfdual Ashtekar connection $\ou{A}{A}{B}$ to an $SU(2)_n$ connection---the $SU(2)_n$ Ashtekar--Barbero connection \cite{Barberoparam,Immirziparam}. The mechanism behind this reduction is easy to understand: Consider first the symplectic term of the discretized BF action for just the $\pi$- and $\omega$-spinors:
\begin{equation}
\oint_{\partial f}\pi_A D\omega^A+\CC=\oint_{\partial f}\di t\Big(\pi_A \frac{\di}{\di t}\omega^A+\pi_A\ou{\Lambda}{A}{B}\omega^B\Big)+\CC,\label{AB1}
\end{equation}
where the Lagrange multiplier $\ou{\Lambda}{A}{B}\in\mathfrak{sl}(2,\C)$ denotes the selfdual connection contracted with the tangent vector $\dot{\alpha}(t)\in T_{\alpha(t)}f$ of the loop $\alpha=\partial f$ bounding the underlying face: $\ou{\Lambda(t)}{A}{B}=\ou{[A_{\alpha(t)}(\dot\alpha)]}{A}{B}$. We now want to impose the simplicity constraints \eref{KLsimpl}, i.e. $\beta L_\alpha+K_\alpha=0$, and see what happens to the symplectic potential. The generators $L_\alpha$ and $K_\alpha$ (as defined in \eref{LKdef}) are however not yet visible in the action \eref{AB1}, but appear only if we decompose $\ou{\Lambda}{A}{B}$ into boosts ($\mathrm{K}^\mu$) and rotations ($\Gamma^\mu$):
\begin{equation}
\ou{\Lambda}{A}{B}=\big(\Gamma^\mu+\I \mathrm{K}^\mu\big)\ou{\tau}{A}{B\mu},\label{AB2}
\end{equation}
where both components are real and purely spatial: $\Gamma^\mu n_\mu=0=\mathrm{K}^\mu n_\mu$, while $\ou{\tau}{A}{B\alpha}$ are the Pauli matrices \eref{Pmatrx}. We can then insert the decomposition \eref{AB2} back into the action \eref{AB1}, and write the generators $L_\alpha$ and $K_\alpha$  as the real and imaginary parts of the selfdual flux $-2\Pi_{AB}=\pi_{(A}\omega_{B)}$:
\begin{equation}
L_\alpha= -\omega_A\pi_B\ou{\tau}{AB}{\alpha}+\CC,\qquad
K_\alpha=\I\omega_A\pi_B\ou{\tau}{AB}{\alpha}+\CC\label{LKdef2}
\end{equation}
Finally, we use the simplicity constraints $\beta L_\alpha+K_\alpha=0$ and bring the face action \eref{AB1} into the following form:
\begin{equation}
\oint_{\partial f}\pi_A D\omega^A+\CC\stackrel{\text{cons}}{=}\oint_{\partial f}\di t\Big(\pi_A \frac{\di}{\di t}\omega^A+\bar\pi_{A'} \frac{\di}{\di t}\bar{\omega}^{A'}+\big(\Gamma^\mu+\beta \mathrm{K}^\mu\big)L_\mu\Big),
\label{AB3}
\end{equation}
where $\stackrel{\text{cons}}{=}$ means equality up to terms constrained to vanish.
Equation \eref{AB3} suggests to introduce the $SU(2)_n$ Ashtekar--Barbero connection $\mathcal{A}^\alpha:=\Gamma^\alpha+\beta\mathrm{K}^\alpha$ along the underlying edge, and define the corresponding covariant $t$-derivative:
\begin{equation}
\frac{\mathcal{D}}{\di t}
 \begin{pmatrix}
  \omega^A\\
  \bar\pi_{A'}
 \end{pmatrix} =
 \frac{\di}{\di t}
  \begin{pmatrix}
  {\omega}^A\\
  {\bar\pi}_{A'}
 \end{pmatrix} +
  \begin{pmatrix}
   \mathcal{A}^\alpha\ou{\tau}{A}{B\alpha}\omega^B&\\
  - \mathcal{A}^\alpha\ou{\bar\tau}{B'}{A'\alpha}\bar\pi_{B'}&
 \end{pmatrix},\quad\text{with:}\quad
 \mathcal{A}^\alpha:=\Gamma^\alpha+\beta\mathrm{K}^\alpha.\label{ABdef}
\end{equation}
We can thus write the kinetic term of the action as the integral:
\begin{equation}
\oint_{\partial f}\pi_A D\omega^A+\CC\stackrel{\text{cons}}{=}\oint_{\partial f}\pi_A \mathcal{D}\omega^A+\CC\label{AB4}
\end{equation}
Here, $\mathcal{D}$ is the covariant differential with respect to the Ashtekar--Barbero connection $\mathcal{A}^\alpha=\Gamma^\alpha+\beta\mathrm{K}^\alpha$ on the edges. The definition of this connection requires a time-normal---the tetrahedral normals $n^\alpha_e$---attached to the boundary $\alpha=\partial f$ of the underlying face: The boundary $\partial f$ consist of several edges $e_1, e_2, \dots$ (see figure \ref{fig1} for an illustration), each of which represents a dual tetrahedron $T_i$ with a time normal $n^\alpha_{e_i}(t)$ parametrically depending on $t$. As we move forward in $t$, we go from one Lorentz frame to another, and $n^\alpha_{e_i}(t)$ is, therefore, the local representative of the $i$-th tetrahedron's normal as seen from the Lorentz frame at the point $\alpha(t)\in e_i\subset\partial f$ in the boundary of the underlying face.
%-------------------------------------------------------------------------------%  
\subsection{Putting the pieces together---defining the action}\noindent
%-------------------------------------------------------------------------------%
%The next step is to add the linear simplicity constraint \eref{linsimpl} to the spinorial action \eref{spinBF2}. 
The last section gave us two immediate consequences of the simplicity constraints added to the action: (i) the reduction of the $SL(2,\C)$ connection $\ou{A}{A}{B}$ to the $SU(2)_n$ Ashtekar--Barbero connection $\mathcal{A}^\alpha$, and (ii) the appearance of the time-normals $n^\alpha_{e}$ of the elementary tetrahedra as additional configuration variables in the action. With these two facts in mind, we are now ready to develop our proposal for a first-order action for discretized gravity. 

\subparagraph*{Face action} First of all, we look at the contributions from the individual faces. We take the discretized topological BF action on a face, i.e. equation \eref{spinBF2}, introduce additional Lagrange multipliers and add the discretized simplicity constraints \eref{linsimpl}. We have, however, studied several functionally equivalent forms of those constraints: The three alternatives (\ref{linsimpl}, \ref{KLsimpl}), and \eref{spinsimpl} are all equally good. Which one of them should we add to the action? We are working in the spinorial representation, so the simplicity constraints in the form of $V=0=W=\bar{W}$ \eref{spinsimpl} are the preferred choice for us. This choice has however yet another advantage: The system of equations \eref{spinsimpl} has a Lorentz invariant component $V=\I/(\beta+\I)\pi_{A}\omega^{A}+\CC$, and cleanly separates it from those other parts $W=n^{AA'}\bar\pi_{A}\omega_{A'}$ and $\bar W$ that are invariant only under the little group $SU(2)_{n}\subset SL(2,\C)$. If we now also remember that the simplicity constraints reduce the $SL(2,\C)$ connection $\ou{A}{A}{B}$ down to the $SU(2)_{n}$ Ashtekar--Barbero connection $\mathcal{A}^{\alpha}$, and replace, therefore, the $SL(2,\C)$ covariant derivative $D$ by the differential $\mathcal{D}$ of the Ashtakar--Barbero connection (shown as as in \eref{AB4}), we are finally led to the following expression:
\begin{align}\nonumber
S_{\mathrm{face}}[Z,\utilde{Z}|\zeta,z,\lambda|\mathcal{A},n]=&
\oint_{\partial f}\Big(
\pi_A\mathcal{D}\omega^A-\utilde{\pi}_A\di\utilde{\omega}^A
-\zeta\big(\utilde{\pi}_A\utilde{\omega}^A-\pi_A\omega^A\big)+\\
&\qquad\quad-\frac{\lambda}{2}\Big(\frac{\I}{\beta+\I}\pi_A\omega^A+\CC\Big)
-z\,n^{AA'}\pi_A\bar{\omega}_{A'}
\Big)+\CC\label{faceactn}
\end{align}
This will be the face contribution to our proposal for a discretized gravitational action. Equation \eref{faceactn} defines an action that depends on seven elements: $Z$ and $\utilde{Z}$ are the twistors parametrizing the discretized holonomy-flux variables (\ref{fluxpar}, \ref{holpar}), while $\zeta\in\C$, $z\in\C$ and $\lambda\in\R$ are additional Lagrange multipliers imposing the area-matching \eref{areamatch} and simplicity constraints \eref{spinsimpl} respectively. The action is also a functional of both the time-normal $n^\alpha$ and the Ashtekar--Barbero connection $\mathcal{A}^{\alpha}$ hiding in the covariant derivative $\mathcal{D}\pi^{A}=\di\pi^{A}+\mathcal{A}^{\alpha}\ou{\tau}{A}{B\alpha}\pi^B$, where, however, also $\ou{\tau}{A}{B\alpha}$ implicitly depends on $n^\alpha$ through \eref{Pmatrx}.

\subparagraph*{Edge action}  The next point concerns the future-oriented time-normals $n^{\alpha}_{e}$ of the individual tetrahedra. The action \eref{faceactn} contains them as additional configuration variables $n^{\alpha}_{e}(t)$ parametrically depending on $t$, the coordinate along the underlying edges. There is, however, no term in the action that would determine this $t$-dependence. This is a problem for us, and we now have to find a way to make the normals proper dynamical variables.

Any proposal for the dynamics of the time-normals ought to respect their geometric interpretation: The vectors $n^{\alpha}_{e}$ are not arbitrary. If we weight them by the volume $\mathrm{Vol}_{e}$ of the corresponding tetrahedron, these normals close at the vertices of the discretization: A four-simplex contains five tetrahedra $T_{i}:i=1,\dots 5$; $\mathrm{Vol}_{i}$ denotes their volume, and $N_{i}^{\alpha}$ be their outwardly oriented time-normal, with $N_i^\alpha=\pm n_i^\alpha$ depending on whether $N^\alpha_i$ is future (past) oriented. These volume-weighted four-normals $N^{\alpha}_{i}\mathrm{Vol}_{i}$ have to sum up to zero, otherwise the tetrahedra would not match together.\footnote{The proof is immediate and follows from Stoke's theorem: We can map the four-simplex into a local Minkowski frame with affine coordinates $X^{\alpha}$, and can thus write the outwardly oriented volume-weighted normal of the $i$-th tetrahedron as the integral: $N_{\alpha}^{i}\mathrm{Vol}_{i}=-\frac{1}{3!}\int_{T_i}\epsilon_{\alpha\beta\mu\nu}\di X^\beta\wedge\di X^\mu\wedge\di X^\nu$. The integrand is exact, hence $\sum_{i=1}^5N^{i}_{\alpha}\mathrm{Vol}_{i}=0$.} The reverse is also true: If $N^{\alpha}_{i}\in\R^{4}$ for $i=1,\dots,5$ are five normalised timelike vectors, while $\mathrm{Vol}_{i}\in\R_{>}$ shall denote the volume of the corresponding tetrahedra, then there exists a unique (up to translations) four-simplex bounded by five spacelike tetrahedra $T_{i}\subset\R^{4}$ with three-volume $\mathrm{Vol}_i$, and outwardly oriented time-normals $N^{\alpha}_{i}$, if and only if:
\begin{equation}
\sum_{i=1}^5N_{\alpha}^{i}\mathrm{Vol}_{i}=0.\label{fourclos}
\end{equation}
What we have described here is essentially the Minkowski theorem generalized to Minkowski space \cite{Minkowskitheorem, polyhdr, wwdiss}. Equation \eref{fourclos} suggests to define for any tetrahedron (dual to an edge $e$) the following four-vector:
\begin{align}
p^\alpha_e:=n^\alpha_e\,\mathrm{Vol}(e),\label{momdef}
\end{align}
where, $n^{\alpha}_{e}$ is the tetrahedron's $T_e$ future-oriented time-normal in the frame over the dual edge $e$, while $\mathrm{Vol}(e)$ denotes the tetrahedron's three-volume. The three-volume is itself a functional of the spinors, implicitly given by the cross-product \cite{Barbieri:1997ks,polyhdr,Rovelliarea}:
\begin{equation}
\mathrm{Vol}^2(e)=\frac{2}{9}\frac{\beta^3\ellp^{6}}{\hbar^{3}}\Big|\epsilon^{\mu\nu\alpha\beta}n^e_\mu L_\nu^{ef_1}L_\alpha^{ef_2}L_\beta^{ef_3}\Big|,\label{voldef}
\end{equation}
where $L_{\alpha}^{ef}$ (\ref{fluxspin}, \ref{LKdef}, \ref{LKdef2}) denotes the rotational part of the momentum flux $\int_{\tau_{f}}\Pi_{\alpha\beta}=\frac{\hbar}{(2\ellp^{2}}\int_{\tau_{f}}(\ast\Sigma_{\alpha\beta}-\beta^{-1}\Sigma_{\alpha\beta})$ through the triangle $\tau_{f}$ dual to the face $f$ in the frame of the tetrahedron $T_e:\partial T_e\supset \tau_{f}$ dual to the edge $e\subset\partial f$.

We now need an additional term in the action that gives us the $t$-evolution of the momentum variables $p^{\alpha}(e)$ along the edges $e(t)$, and is consistent with the conservation law \eref{fourclos} at the vertices. We are taking the momenta as our fundamental variables, so we certainly need a constraint imposing \eref{momdef}, and thus introduce the mass shell condition:
\begin{equation}
C:=\frac{1}{2}\Big(p_\alpha p^\alpha+\mathrm{Vol}^2(e)\Big)\stackrel{!}{=}0.\label{msscons}
\end{equation}
Next, we also need a kinetic term, which should be compatible with the symmetries of the action \eref{faceactn}. One immediate possibility would be to use the covariant derivative of the original $SL(2,\C)$ connection and add the term $X_{\alpha}\frac{D}{\di t}p^{\alpha}$ to the action. %, where the \qq{position} variables $X^{\alpha}$ would play the role of Lagrange multipliers enforcing the momenta to be parallel along the edges. 
We have, however, already seen that the simplicity constraint reduce the local symmetry group to the little group $SU(2)_{n}\subset SL(2,\C)$, thus replacing the selfdual connection $\ou{A}{A}{B}$ by the Ashtekar--Barbero connection $\mathcal{A}^{\alpha}$. The term $X_{\alpha}\frac{D}{\di t}p^{\alpha}$ is therefore no option for us, because the addition of the simplicity constraints has just washed away the original $SL(2,\C)$ connection. A term $X_{\alpha}\dot{p}^{\alpha}$, on the other hand, is still invariant under the little group $p^{\alpha}\rightarrow\ou{U}{\alpha}{\beta}p^{\beta}$, with  $\ou{U}{A}{B}\in SU(2)_{n}=\{\ou{h}{AA'}{BB'}\subset SL(2,\C):p^{AA'}=\ou{h}{A}{B}\ou{\bar{h}}{A'}{B'}p^{BB'}\}$. The ordinary derivative is therefore still compatible with all the symmetries of the action \eref{faceactn}. A viable candidate for our action is therefore the following: Take the symplectic potential $p_{\alpha}\di X^{\alpha}$ and just add the mass-shell condition $C=0$, explicitly:
\begin{equation}
S_{\mathrm{edge}}[x,p|N,\mathrm{Vol}(e)]=\int_e\Big(p_\alpha\di X^\alpha-\frac{N}{2}\big(p_{\alpha}p^{\alpha}+\mathrm{Vol}^2(e)\big)\Big),\label{edgeactn}
\end{equation}
where the one-form $N$ imposes the constraint \eref{msscons}. Variation with respect to the $X^{\alpha}$-variables determines the $t$-dependence of the momentum variables: All momenta $p^{\alpha}_{e}$ are constant along the corresponding edges, thus introducing a notion of distant parallelism. The $X^{\alpha}$-variation of the action yields a remainder $p_{\alpha}\delta X^\alpha|_{\partial e}=p_{\alpha}\delta X^\alpha|_{v_1}-p_{\alpha}\delta X^\alpha|_{v_o}$ at the two vertices $v_{o}$ and $v_{1}$ bounding the underlying edge $e$. We thus need yet another term in the action, otherwise the action would not be functionally differentiable. For every vertex $v$, we thus add the following boundary term:
\begin{equation}
%S_{\mathrm{vertex}}[X,\varPi_{\mathrm{out}},\varPi_{\mathrm{in}}]=
S_{\mathrm{vertex}}[Y_v,\{X_{ev}\}_{e\ni v},\{v_{ev}\}_{e\ni v}]=\sum_{e:e\ni v}\big(Y^{\alpha}_v-X^{\alpha}_{ev}\big)v_{\alpha}^{ev}.\label{vertxactn}
%\Big(\sum_{p^{\alpha}\in\varPi_{\mathrm{out}}} p^\alpha-\sum_{p^\alpha\in\varPi_{\mathrm{in}}}p^{\alpha}\Big)=0,\label{vactn}
\end{equation}
The sum goes over all edges meeting at the vertex $v$. The variation of the Lagrange multipliers $v^e_\alpha$ tell us that the edges meet in a point, i.e. $Y^\alpha-X^\alpha_{ev}=0$ (with $X^\alpha_{ev}$ denoting the value of $X^\alpha_e$ at the bounding vertex $v\in\partial e$). The variation of $X^\alpha_{ev}$, on the other hand, cancels the boundary term coming from the edge action \eref{edgeactn}, and yields $v^\alpha_{ev}=\pm p^\alpha_{ev}$ depending on the orientation of the underlying edge. Finally, there is the variation of the Lagrange multiplier $Y^\alpha_v$ yielding the conservation law $\sum_{e\ni v}v^\alpha_{ev}=0$ for the momenta $p^\alpha_{ev}=\pm v^\alpha_{ev}$, i.e. the four-dimensional closure constraint \eref{fourclos}.

%where $X^{\alpha}$ is the corresponding Lagrange multiplier, while $\varPi_{\mathrm{out}}$ ($\varPi_{\mathrm{in}}$) denotes the set of volume-weighted normals for all outgoing (incoming) tetrahedra (a tetrahedron in a four-simplex is outgoing (incoming) if its outwardly oriented four-normal is future (past) pointing).

The momenta $p^{\alpha}_{e}(t)$ are local sections of a $\R^4$-vector bundle over the edges of the discretization. For the position variables $X^{\alpha}_{e}$, we take a slightly different target space: $X^\alpha_e(t)$ is not an element of a vector space, the standard fibre is the affine Minkowski space $\mathbb{M}^4$ itself. We expect that the corresponding affine fibre bundle will generally have a non-trivial topology. Over every edge of the discretization the corresponding affine gauge connection is flat, curvature manifests itself only if we go once around a loop. In fact, the equations of motion for $X^\alpha_e(t)$ define the affine parallel transport of $X^\alpha_e(0)$ along the underlying edge $e$. The transition functions between different local trivializations are translations, their sum around a loop gives us the affine curvature in a face as a dislocation vector $b^\alpha_f\in\R^4$. The dynamics of the $(X^\alpha,p_\alpha)$-variables is therefore just a version of locally flat teleparallelism \cite{Hehl:2007bn, Obukhov:2003aa,Pereira:2002ff,Gronwald:1995mb,Caselle:1989hd}. 

\subparagraph*{Putting the pieces together} We are now ready to close this section, summarize our proposal, and give the definition of the action. We have started with a four-dimensional orientable spacetime manifold $M$, and introduced a 
simplicial decomposition to discretize the topological BF action as a sum over the elementary faces of the discretization \eref{spinBF2}. Next, we studied the simplicity constraints imposing the geometricity of the elementary tetrahedra. This led us to a proposal for a constrained action on the simplicial faces \eref{faceactn}. The simplicity constraints contain the tetrahedral time-normals as additional configuration variables. We argued that a robust theory should also treat these time-normals as dynamical variables. This requires additional terms in the action. We gave a proposal for such an action consisting of two parts. The first part \eref{edgeactn} is an additional contribution from each edge in the discretization, the other part \eref{vertxactn} is a boundary term at the simplicial vertices. Putting everything together---the BF action on the faces, the simplicity constraints, the edge-action determining the $t$-evolution of the tetrahedral normals, and the boundary term at the vertices---we are thus led to the following action for a theory of first-order Regge gravity:
\begin{align}\nonumber
S_{\text{spin-Regge}}=&\sum_{f:\text{faces}}S_{\text{face}}\big[Z_f,\utilde{Z}_f\big|\zeta_f,z_f,\lambda_f\big|\mathcal{A}_{\partial f},n_{\partial f}\big]+
\sum_{e:\text{edges}}S_{\text{edges}}\big[X_e,p_e\big|N_e,\mathrm{Vol}(e)\big]+\\
&+\sum_{v:\text{vertices}}S_{\mathrm{vertex}}\big[Y_v,\{X_{ev}\}_{e\ni v},\{v_{ev}\}_{e\ni v}\big].\label{spinactn}
\end{align}
The action depends on several variables: The twistors $Z_f$ and $\utilde{Z}_f$ parametrize the original holonomy-flux variables (\ref{fluxpar}, \ref{holpar}) and define a map $Z_f:\partial f\rightarrow\T, t\mapsto (\bar\pi_{A'}(t),\omega^A(t))$ into twistor space $\T\simeq\C^4$, the Lagrange multipliers $\zeta_f(t)\in\C$, $z_f(t)\in\C$ and $\lambda_f(t)\in\R$ impose the area-matching \eref{areamatch} and simplicity constraints \eref{spinsimpl}, $n_{\partial f}:\partial f\mapsto \R^4$ is the time normal along the boundary of the underlying face, and $\mathcal{A}_{\partial f}$ is the corresponding Ashtekar--Barbero connection \eref{ABdef} hiding in the covariant derivative $\mathcal{D}\pi^A_f=\dot{\pi}^A_f+\mathcal{A}^\alpha\ou{\tau}{A}{B\alpha}\pi^B$; next there are the other two terms: $p_e$ is the volume weighted time-normal along an edge $e\subset \partial f$ bounding a face $f$, $X_e$ is its conjugate momentum and $N_e$ is a Lagrange multiplier imposing the mass shell condition $p_\alpha^ep^\alpha_e=-\mathrm{Vol}^2(e)$, with $\mathrm{Vol}(e)$ denoting the volume \eref{voldef} of the dual tetrahedron; $X_{ev}$, on the other hand, is the value of $X_e$ at a bounding vertex $v\in\partial e$, while, finally, $Y_v$ and $v_{ev}$ are additional Lagrange multipliers, needed to glue the individual edges meeting at a vertex $v$.

Equation \eref{spinactn} is our final proposal, a proposal for an action for discretized gravity in first-order variables. At this point, neither do we know of any global solutions of the resulting equations of motion on an arbitrary two-complex, nor do we have a proof that they would correspond to any physical spacetime geometry. Yet, we do have positive evidence in favor of our proposal. First of all, we will see, that the constraint algebra closes, and that there are no secondary constraints. Then, the model has curvature. This curvature lies in the faces dual to the elementary triangles, and is given, just as in Regge calculus, by the sum over the boost-angles between the adjacent tetrahedra. Finally, and most importantly, the solutions of the equations of motion have a geometric interpretation and define a twisted geometry \cite{freidelsimotwist, twist, twistedconn}. Twisted geometries are discrete geometries found in the semi-classical limit of loop quantum gravity \cite{rovelli, thiemann, status, zakolec}. They are similar to Regge geometries insofar as they represent a collection of flat tetrahedra glued along their bounding triangles, but unlike Regge geometries there are no unique length variables: Every triangle has a unique area, and every tetrahedron has a unique volume, but the length of a triangle's bounding side exists only locally. %Passing from one vertex to the next, the tetrahedra undergo a shear.%, and this shear contributes to the spacetime curvature.

%-------------------------------------------------------------------------------%
\section{Dynamics of the theory}\label{secII}\noindent
%-------------------------------------------------------------------------------%
In the last section, we gave a proposal for a gravitational action on a simplicial lattice. Now it is time to study the dynamics. The action \eref{spinactn} is local in $t$, and so are the resulting equations of motion, that tell us how the elementary configuration variables change as we move forward in $t$ and go from one vertex to the next. This $t$-variable does however not have an immediate physical interpretation. It is no physical time, and does not measure duration as given by a clock.
\subsection{Hamiltonian formulation}
\noindent\subparagraph*{Symplectic structure}  Let us first fix an arbitrary edge $e$ in the discretization. Restricting our analysis to just a single edge is a matter of convenience. It allows us to use a condensed notation and drop rather annoying edge-labels. We thus write $p^\alpha\equiv p^\alpha_e$, and equally for all other variables when useful. All our results translate immediately to any other edge. Now, the edge $e$ is adjacent to four bounding faces $f_1,\dots,f_4$: $\forall i:e\subset\partial f_i$, because any edge is dual to a tetrahedron containing \emph{four} triangles $\tau_i$, each of which is dual to a face $f_i$. Each of these faces carries an orientation. For a given edge $e$, we can now always choose this orientation such that the induced orientation of $\partial f_i$ agrees with the orientation of $e$. That both $\partial f_i$, and $e\subset \partial f_i$ have the same orientation is possible only locally. In general, we would have to introduce additional sign factors $\varepsilon(e,f)=\pm 1$ taking care of the relative orientation between $f$ and $e$. Remember, this we have already done when studying the Gauß law in equation \eref{BFGauss} above. Here, we are looking only at a single edge, and can therefore assume that $e$ and $f$ have the same orientation, hence $\forall i:\varepsilon(e,f_i)=1$ without loss of generality.

Going back to our definition of the action \eref{spinactn}, and looking at the contributions from edges and adjacent faces, i.e. equations \eref{edgeactn} and \eref{faceactn}, we can see that the action $S_{\text{spin-Regge}}$ has the structure of a general covariant one-dimensional constrained system \cite{rovelli}, i.e. a system described by an action of the form $S=\int \di t (P_i\dot{Q}^i- \mu^IC_I(P,Q))$ with canonical coordinates $P_i$ and $Q^j$: $\{P_i,Q^j\}=\delta^j_i$, and Lagrange multipliers $\mu^I$ imposing the constraints $C_I(P,Q)=0$. The first terms in the action (\ref{faceactn}, \ref{edgeactn}) define the symplectic structure. The only non-vanishing Poisson brackets are, in fact:
\begin{subequations}
\begin{align}
&\phantom{\{}\big\{p_\alpha,X^\beta\big\}=\delta^\beta_\alpha,\phantom{\}}\label{sympstruct1}\\
&\left.\begin{aligned}
&\big\{\pi_A^i,\omega^B_j\big\}=+\delta^i_j\delta^B_A,\qquad
\big\{\bar\pi_{A'}^i,\bar\omega^{B'}_j\big\}=+\delta^i_j\delta^{B'}_{A'},\\
&\big\{\utilde{\pi}_A^i,\utilde{\omega}^B_j\big\}=-\delta^i_j\delta^B_A,\qquad
\big\{\utilde{\bar\pi}_{A'}^i,\utilde{\bar\omega}^{B'}_j\big\}=-\delta^i_j\delta^{B'}_{A'},
\end{aligned}\quad\right\}\label{sympstruct2}
\end{align}
\end{subequations}
where e.g. $\omega^A_i$ is a shorthand notation for the spinor field $\omega^A_{f_i}:\partial f_i\rightarrow\C^2$ along the boundary of the face $f_i$.
\subparagraph*{Hamilton equations} The action \eref{spinactn} has already a Hamiltonian form: $S=\int \di t (P_i\dot{Q}^i- \mu^IC_I(P,Q))$. The evolution equations are therefore generated by the Hamiltonian $H=\mu^IC_i$ as: $\dot{P}_i=\{H,P_i\}$ and $\dot{Q}^i=\{H,Q^i\}$. Going back to our action (\ref{faceactn}, \ref{edgeactn}, \ref{spinactn}), we can thus immediately read off the Hamiltonian: 
\begin{equation}
H=\mathcal{A}^\alpha G_\alpha+\sum_{i=1}^4
\Big(\zeta^i\varDelta_i+\bar{\zeta}^i\bar{\varDelta}_i+z^iW_i+\bar{z}^i\bar{W}_i+\lambda^iV_i\Big)+NC,\label{Ham}
\end{equation}
where $\mathcal{A}^\alpha$ is the Ashtekar--Barbero connection along the edge \eref{ABdef}, and $\zeta^i\in\C$, $z_i\in \C$ and $\lambda^i\in\R$ are additional Lagrange multipliers imposing the constraints of the theory. The Hamiltonian generates the $t$-evolution along the edges of the discretization, e.g.:
\begin{equation}
\frac{\di}{\di t}\omega^A_i=\big\{H,\omega^A_i\big\}.\label{hamflow}
\end{equation}
The invariance of the action under reparametrizations of the underlying edge implies the vanishing of the Hamiltonian: $G_\alpha$ is the rotational part of the Gauß constraint, $\varDelta_i$ is the area-matching constraint reducing spinors to the original holonomy-flux variables, $W_i$ and $V_i$ are the spinorial equivalent of the simplicity constraints, while $C$ denotes the mass-shell condition. Explicitly:
\begin{subequations}
\begin{align}
G_\alpha&=\sum_{i=1}^4\ou{\tau}{AB}{\alpha}\omega^i_A\pi^i_B+\CC\equiv-\sum_{i=1}^4L_\alpha^i\stackrel{!}{=}0,\\
%\end{align}
%and furthermore:
%\begin{align}
\varDelta_i&=\utilde{\pi}_A^i\utilde{\omega}^A_i-\pi_A^i\omega^A_i\stackrel{!}{=}0,\\
W_i&=n^{AA'}\pi_A^i\bar{\omega}_{A'}^i\stackrel{!}{=}0,\\
V_i&=\frac{\I}{\beta+\I}\pi_A^i\omega^A_i+\CC\stackrel{!}{=}0,\\
C&=\frac{1}{2}\Big(p_\alpha p^\alpha+\mathrm{Vol}^2\Big)\stackrel{!}{=}0.\label{mssshell}
\end{align}\label{consset}
\end{subequations}
Here $p_\alpha$ is the tetrahedron's volume-weighted normal \eref{momdef}, while $\mathrm{Vol}^2\propto\frac{2}{9}\epsilon^{\alpha\beta\mu}L_\alpha^1L_\beta^2L_\mu^3$ is the squared volume of the tetrahedron, $L_\alpha^i$ is the rotational part \eref{LKdef} of the momentum flux through the tetrahedron's $i$-th triangle, $\ou{\tau}{AB}{\alpha}$ are the Pauli matrices \eref{Pmatrx} with respect to the tetrahedron's time-normal $n^\alpha=n^{AA'}:p^\alpha=n^\alpha\mathrm{Vol}$ and $\beta>0$ is again the Barbero--Immirzi parameter. Notice also, that the summation convention applies only to the spinor indices $A,B,C,\dots$; we do not sum over paired indices $i,j,k,\dots$ labeling the bounding triangles, hence: $\pi_A^i\omega^A_i\neq \sum_{i=1}^4\pi_A^i\omega^A_i$.

\subparagraph*{Constraint algebra and Dirac analysis} The next step is to study the constraint algebra. We will address the following questions: Does the Hamiltonian flow \eref{hamflow} preserve the system of constraints \eref{consset}? Are there any secondary constraints? Do we have both first- and second-class constraints? And if there are any first-class constraints, what kind of symmetries do they generate? And if we also get second-class constraints, how does this affect the Lagrange multipliers $N,\lambda_i\in\R$ and $z_i,\zeta_i\in\C$ in the action?

Let us start our analysis with the Gauß constraint $G_\alpha$. %To calculate the relevant Poisson brackets we need the symplectic structure (\ref{sympstruct1}, \ref{sympstruct2}) together with the properties of the Pauli matrices $\ou{\tau}{A}{B\alpha}$ (\ref{Pmatrx}, \ref{Paulident}). 
In fact, $G_\alpha$ Poisson-commutes with all other constraints: The anti-Hermiticity of $\ou{\tau}{A}{B\alpha}$ with respect to the $SU(2)_n$-norm $\delta_{AA'}=\sigma_{AA'\alpha}n^\alpha$ implies that the Gauß constraint $G_\alpha$ generates $SU(2)_n$ transformations, and $G_\alpha$ commutes, therefore, with all $SU(2)_n$ invariants. Now, all the other constraints $\varDelta_i$, $C$, $W_i$ and $V_i$ are just linear combinations of the $SU(2)_n$-invariant contractions $E_i=\pi_A^i\omega^A_i$, $\|\omega_i\|^2_n=\delta_{AA'}\omega^A_i\bar\omega^{A'}_i$ and $\langle\omega_i,\pi_i\rangle=\delta_{AA'}\bar{\omega}^{A'}_i\pi^A_i$, hence:
\begin{equation}
\big\{G_\alpha,\varDelta_i\big\}=\big\{G_\alpha,\bar{\varDelta}_i\big\}=\big\{G_\alpha,V_i\big\}=
\big\{G_\alpha,W_i\big\}=\big\{G_\alpha,\bar{W}_i\big\}=\big\{G_\alpha,C\big\}=0.\label{Gzero}
\end{equation}
There is one Poisson bracket missing in this list. In fact, the $\mathfrak{su}(2)_n$ commutation relations \eref{Paulident} among the Pauli matrices $[\tau_\alpha,\tau_\beta]=\uo{\epsilon}{\alpha\beta}{\mu}\tau_\mu$ immediately translate into commutation relations for the Gauß constraint $G_\alpha=-\sum_{i=1}^4L_\alpha^i=\sum_{i=1}^4\ou{\tau}{AB}{\alpha}\pi_A^i\omega_B^i+\CC$ itself:
\begin{equation}
\big\{G_\alpha,G_\beta\big\}=\uo{\epsilon}{\alpha\beta}{\mu}G_\mu,\label{rotalg}
\end{equation}
where $\epsilon_{\alpha\beta\mu}=n^\nu\epsilon_{\nu\alpha\beta\mu}$ denotes the three-dimensional internal Levi-Civita tensor with respect to the tetrahedron's time-normal $n^\alpha\propto p^\alpha$. 
%To prove \eref{Gzero} from the elementary Poisson brackets \eref{sympstruct2}, it is very useful to employ the generalized Pauli identity \eref{Paulident}: $\tau_\alpha\tau_\beta=-\frac{1}{4}h_{\alpha\beta}\mathds{1}+\frac{1}{2}\uo{\epsilon}{\alpha\beta}{\mu}\tau_\mu$, with $h_{\alpha\beta}=n_\alpha n_\beta+\eta_{\alpha\beta}$.

For the remaining Poisson brackets between $C$, $\varDelta_i$, $W_i$ and $V_i$ let us only give those, that do not vanish identically. They are the following:
\begin{align}
&\begin{aligned}
&\big\{\varDelta_i,W_j\big\}=\delta_{ij}W_i,\\
&\big\{\bar{\varDelta}_i,\bar{W}_j\big\}=\delta_{ij}\bar{W}_i,\\
&\big\{V_i,W_j\big\}=-\frac{2\I\beta}{\beta^2+1}\delta_{ij}W_j,
\end{aligned}\quad\left.
\begin{aligned}
&\big\{\varDelta_i,\bar{W}_j\big\}=-\delta_{ij}\bar{W}_i,\\
&\big\{\bar{\varDelta}_i,{W}_j\big\}=-\delta_{ij}{W}_i,\\
&\big\{V_i,\bar{W}_j\big\}=+\frac{2\I\beta}{\beta^2+1}\delta_{ij}\bar{W}_j,
\end{aligned}\quad
\right\}\label{fclass}\\
&\,\big\{W_i,\bar{W}_j\big\}=-\I\delta_{ij}\mathfrak{Im}(E_i)=-\frac{1}{2}\delta_{ij}\big(\pi_A^i\omega^A_i-\CC\big).\label{sclass}
%\big\{,\big\}
\end{align} 
Notice the absence of the mass-shell condition $C=0$ in this list. Indeed it Poisson-commutes with all other constraints: Consider first the generator $L_\alpha^i=-\omega^A_i\pi^B_i\ou{\tau}{AB}{\alpha}+\CC$ of $SU(2)_n$ transformations in the $i$-th triangle:
\begin{equation}
\bigg\{L_\alpha^i,\left(\begin{aligned}&\bar{\pi}^j_{A'}\\&\omega^A_i\end{aligned}\right)\bigg\}=
\delta_{ij}\left(\begin{aligned}-&\ou{\bar\tau}{B'}{A'\alpha}\bar{\pi}^j_{B'}\\&\ou{\tau}{A}{B\alpha}\omega^B_i\end{aligned}\right).
\end{equation}
This generator trivially commutes with the $SU(2)_n$ invariant contractions $\pi_A^i\omega^A_i$ and $n_{AA'}\pi^A_i\bar{\omega}^{A'}_i$, and it thus also commutes with the constraints $\varDelta_i$, $V_i$ and $W_i$ themselves.
Now, $L_\alpha^i$ appears linearly in the squared volume $\mathrm{Vol}^2\propto\frac{2}{9}\epsilon^{\alpha\beta\mu}L_\alpha^iL_\beta^jL_\mu^k$ of the tetrahedron, and equally in the constraint $C$ itself  (\ref{voldef}, \ref{msscons}). Therefore: $\{C,\varDelta_i\}=\{C,\bar{\varDelta}_i\}=\{C,V_i\}=\{C,W_i\}=\{C,\bar{W}_i\}=0$. The volume functional $\mathrm{Vol}\propto\frac{2}{9}\epsilon^{\alpha\beta\mu}L^i_\alpha L^j_\beta L^k_\mu$ is itself an $SU(2)_n$-invariant contraction of the fluxes, hence also $\{G_\alpha,C\}=0$.

Looking back at the fundamental Poisson brackets, i.e. equations \eref{sclass}, \eref{fclass} and \eref{rotalg} we can now immediately identify the list of first- and second-class constraints: $\{G_\alpha, C, \varDelta_i, V_i:i=1,\dots,4\}$ are of first class, while the four remaining constraints $W_i$ for $i=1,\dots,4$ are of second class. The latter is a direct consequence of the fact that the right hand side of equation \eref{sclass} does not vanish weakly, i.e.: $\{W,\bar{W}\}=\I\Im(\pi_A\omega^A)\neq 0$. Indeed, the Lorentz invariant contraction $\pi_A\omega^A$ is proportional\footnote{\label{arfoot}We can derive this proportionality as follows. First of all, in a locally flat geometry, the triangle's area is given by the square root of the corresponding fluxes: $\mathrm{Ar}=\sqrt{\frac{1}{2}|\Sigma_{\alpha\beta}\Sigma^{\alpha\beta}|}$. Now, $\Sigma_{\alpha\beta}\Sigma^{\alpha\beta}=2\Sigma_{AB}\Sigma^{AB}+\CC$ Next, $\Sigma_{AB}=\frac{2\ellp^2}{\I\hbar}\frac{\beta}{\beta+\I}\Pi_{AB}$ (see equation \eref{pisigma}). If we now also employ the spinorial parametrization $\Pi_{AB}=-2\pi_{(A}\omega_{B)}$ together with the simplicity constraint $V=\I/(\beta+\I)\pi_A\omega^A+\CC=0$, we eventually get the desired result $\mathrm{Ar}=\hbar^{-1}\beta\ellp^2/(\beta+\I)\,\pi_A\omega^A$.} to the triangle's area $\mathrm{Ar}$, the precise relation being: $\beta\ellp^2\pi_A\omega^A=\hbar(\beta+\I)\mathrm{Ar}$, which vanishes only if the triangle is either null, or shrunken to a line or point. We have already excluded both cases previously, so $\{W,\bar{W}\}\neq 0$ for all relevant cases. So, indeed, $\{W_i:i=1,\dots,4\}$ defines the set of second-class constraints.

Are these already all constraints, or do we need any additional secondary constraints? We don't, and this can be seen as follows: First of all, the Hamiltonian \eref{Ham} is itself a sum over constraints, and therefore trivially commutes with our set of first-class constraints:
\begin{equation}
\frac{\di}{\di t}
\begin{pmatrix}
C&\varDelta_i&G_\alpha&V_i
\end{pmatrix}=
\Big\{H,
\begin{pmatrix}
C&\varDelta_i&G_\alpha&V_i
\end{pmatrix}
\Big\}\stackrel{\text{cons}}{=}0,
\end{equation}
where $\stackrel{\text{cons}}{=}$ denotes equality up to terms constrained to vanish. The only subtlety arises from the second-class constraints. Looking back at their mutual Poisson bracket, i.e. equation \eref{sclass} we get, in fact:
\begin{equation}
\frac{\di}{\di t}W_i=\big\{H,W_i\big\}\stackrel{\text{cons}}{=}\sum_{j=1}^4\bar{z}^j\big\{\bar{W}_j,W_i\big\}=\frac{z^i}{2}\big(\bar{\pi}_{A'}^i\bar{\omega}^{A'}_i-\pi_A^i\omega^A_i\big).\label{Tevolv}
\end{equation}
Notice that the summation convention only holds for the spinorial indices $A,B,\dots$, we do not sum over repeated indices $i,j,\dots\in\{1,\dots,4\}$ labeling the four triangles in the tetrahedron. Now, once again, the right hand side of equation \eref{Tevolv} is proportional to the $i$-th triangle's area (see footnote \ref{arfoot} for an explanation), hence vanishes for generic triangles only for $z^i=0$. We have thus found, that the Hamiltonian preserves the constraints of the system only if we demand that the Lagrange multiplier in front of the second class constraints vanishes. We thus set:
\begin{equation}
\forall i=1,\dots 4:z^i=0
\end{equation}
The physical Hamiltonian is therefore just given by a sum over the first-class constraints of the system:
\begin{equation}
H_{\mathrm{phys}}=\mathcal{A}^\alpha G_\alpha+\sum_{i=1}^4\big(\zeta^i \varDelta_i+\bar{\zeta}^i\bar{\varDelta}_i+\lambda^iV_i\big)+NC,\label{physham}
\end{equation}
which is, of course, what always happens for reparamtrization invariant systems---timeless systems, systems lacking any preferred external clock \cite{rovelli}.

\subsection{Twisted geometries}\noindent
Now that we have the Hamiltonian at hand, we should ask what kind of geometries it generates. We will identify four features any solution of the equations of motion must have: Point (i) shows that the dynamics in the  $(\utilde{\pi},\utilde{\omega})$-spinors is trivial. The $(\utilde{\pi},\utilde{\omega})$-spinors are spectators evolving just as if the theory were topological. The next point, point (ii) concerns the volume-weighted normals $p_\alpha^e=n_\alpha^e\mathrm{Vol}(e)$ and their canonical conjugate variables $X^\alpha_e$. We will show that $X^\alpha_e$ and $p_\alpha^e$ describe a system of massive particles moving in a locally flat ambient space. Every simplicial edge is in one-to-one correspondence with one of these particles: Every particle represents a tetrahedron with its three-volume representing the particle's mass, its volume-weighted normal representing the particle's four-momentum, and its scattering events representing the simplicial vertices. The next point, point (iii) looks at the dynamics in the $(\pi,\omega)$-sector. The $(\pi,\omega)$-spinors have a very interesting dynamics, they evolve so as to generate a \emph{twisted geometry}, the kind of geometry found from the semi-classical limit of loop quantum gravity. The last point, point (iv) briefly summarizes the local gauge symmetries of the model.

\subparagraph*{(i) Dynamics of the tilde-spinors} The $(\utilde{\pi},\utilde{\omega})$-spinors enter the physical Hamiltonian \eref{physham} only through the area matching constraint $\varDelta=\utilde{\pi}_A\utilde{\omega}^A-\pi_A\omega^A$, which just generates overall rescalings of the spinors. Going back to the definition of the symplectic structure \eref{sympstruct2} we get, in fact:
\begin{equation}
\frac{\di}{\di t}
\left(\begin{aligned}&\utilde{\bar\pi}_{A'}^i\\&\utilde{\omega}^A_i\end{aligned}\right)=
\bigg\{H_{\mathrm{phys}},\left(\begin{aligned}&\utilde{\bar\pi}_{A'}^i\\&\utilde{\omega}^A_i\end{aligned}\right)\bigg\}=
\left(
\begin{aligned}
&\bar\zeta^i\{\bar\varDelta_i,\utilde{\bar\pi}_{A'}^i\}\\
&\zeta^i\{\varDelta_i,\utilde{\omega}^A_i\}
\end{aligned}\right)=\left(
\begin{aligned}\phantom{-}&\bar\zeta^i\utilde{\bar\pi}_{A'}^i\\
-&\zeta^i\utilde{\omega}^A_i\end{aligned}\right).\label{tildedyn}
\end{equation}
The evolution of the $(\utilde{\pi},\utilde{\omega})$-spinors is therefore unaffected by the addition of the simplicity constraints $W$ and $V$. The tilde-spinors evolve just as if we were still in the topological theory \eref{spinevolv}. Closing the edges to a loop, and using the periodic boundary conditions\footnote{The spinors are continuous along the boundary of the underlying face, implying periodic boundary conditions if we go once around the loop: $\pi^A_i|_{\alpha_i(0)}=\pi^A_i|_{\alpha_i(1)}$ and equally for $\omega^A_i$, where $\alpha_i(t)$ for $t\in[0,1]$ parametrizes the boundary of the face $f_i$ dual to the $i$-th triangle.} for the spinors, we thus get again a restriction on the integral ${\int_{\partial f_i}}\!\di t\,\zeta^i$ of the Lagrange multiplier over the boundary of the underlying face. This is the same type of quantization condition that we have already found in \eref{BFflata}: 
\begin{equation}
\mathclap{\int_{\partial f_i}}\di t\,\zeta^i=2\pi\I\,n^i,\qquad n^i\in\Z.\label{windnumb}
\end{equation}
At this point we can only speculate about the physical meaning of this winding number $n_f$ attached to each face in the discretization. The Lagrange multiplier $\zeta_f$ is dual to the area matching constraint $\varDelta_f$, which has dimensions of an area. In Regge calculus areas are always dual to angles. It would be, in fact, very tempting to think of $\zeta_f$ as a local Lorentz-angle winding around the dual triangle $\tau_f$, as a boost-angle along the lines of \cite{Bodendorfer:2013hla} with its imaginary part counting the crossings of the lightcone. % Let us stress, however, that 
 At this point there is, however, very little evidence for this relation, and we thus rather resume the main thread of the paper.

\subparagraph*{(ii) Dynamics of the p-X variables} Let us now study the dynamics of the $p$-$X$-variables. For $p^\alpha$ the situation is simple: The Hamiltonian has no explicit dependence on $X^\alpha$, it therefore commutes with $p_\alpha$, and we immediately find:
\begin{equation}
\frac{\di}{\di t}p_\alpha=\big\{H_{\text{phys}},p_\alpha\big\}=0.\label{Pevolv}
\end{equation}
The evolution equations for $X^\alpha$ require a little more patience. The momentum variable $p^\alpha$ enters the Hamiltonian in two terms: First of all, there is the Gauß constraint $G_\alpha=-\sum_{i=1}^{4}L_\alpha^i$ implicitly depending on $p^\alpha\propto n^\alpha$ through the definition of the rotational part $L_\alpha:=\uo{\epsilon}{\alpha\beta}{\mu\nu}\Pi_{\mu\nu}n^\beta$ of the momentum flux $\Pi_{\alpha\beta}$, see (\ref{fluxdef}, \ref{LKdef}). The second contribution comes from the mass-shell condition \eref{msscons} $C=\frac{1}{2}(p_\alpha p^\alpha+\mathrm{Vol}^2)$. Notice a further subtlety: The momentum variable does not only appear in the first term $\propto p_\alpha p^\alpha$, it also enters the tetrahedron's three-volume \eref{voldef}: $\mathrm{Vol}^2\propto\epsilon^{\alpha\beta\mu\nu}n_\alpha L^i_\beta L^j_\mu L^k_\nu$ through both $n_\alpha\propto p_\alpha$ and the definition of the $SU(2)_n$ generators: $L_\alpha:=\uo{\epsilon}{\alpha\beta}{\mu\nu}\Pi_{\mu\nu}n^\beta$. 

To calculate the relevant Poisson brackets $\{H_{\text{phys}},X^\alpha\}=\{\mathcal{A}^\alpha G_\alpha+NC,X^\alpha\}$, let us first study the differential $\delta_n[L_\alpha]$ of the $SU(2)_n$ generator $L_\alpha$ under infinitesimal variations of the underlying time-normal $n^\alpha$ alone. Looking back at the definitions for both $L_\alpha$ and $K_\alpha$, i.e. \eref{LKdef}, we immediately get:  
\begin{equation}
\uo{h}{\alpha}{\beta}\delta_n[L_\beta]=
\uo{h}{\alpha}{\beta}\uo{\epsilon}{\beta\mu}{\rho\sigma}\Pi_{\rho\sigma}\delta n^\mu=
-2\uo{h}{\alpha}{\beta}\uo{\epsilon}{\beta\mu}{\rho\sigma}n_\rho n^\nu\Pi_{\nu\sigma}\delta n^\mu=
\delta n^\mu\uo{\epsilon}{\alpha\mu}{\sigma}K_\sigma,\end{equation}
where $h_{\alpha\beta}=-n_\alpha n_\beta+\eta_{\alpha\beta}$ denotes the spatial metric induced by $n_\alpha$. If we now also employ the simplicity constraints in the form of \eref{KLsimpl}, i.e. $K_\alpha{=}-\beta L_\alpha$, we see, that we can always absorb the $n^\alpha$-variation of $L_\alpha$ into an infinitesimal $SU(2)_n$ transformation: 
\begin{equation}
\uo{h}{\alpha}{\beta}\delta_n[L_\beta]\stackrel{\text{cons}}{=}-\beta
\uo{\epsilon}{\alpha\mu}{\beta}\delta n^\mu L_\beta,\label{fluxvar}\end{equation}
where $\epsilon_{\alpha\mu\nu}=n^\beta\epsilon_{\beta\alpha\mu\nu}$ is the three-dimensional Levi-Civita tensor, and $\stackrel{\text{cons}}{=}$ denotes again equality up to terms constrained to vanish. % Equation \eref{fluxvar} shows us that variations of $n^\alpha$ induce infinitesimal $SU(2)_n$ transformation of $L_\alpha$. 
This further implies that any variation of the time-normal $n^\alpha$ preserves the $SU(2)_n$ Gauß constraint:
\begin{equation}
\delta_n\big[\mathcal{A}^\alpha G_\alpha\big]\stackrel{\text{cons}}{=}-\sum_{i=1}^{4}\mathcal{A}^\alpha\delta_n\big[L_\alpha^i\big]\stackrel{\text{cons}}{=}\beta\mathcal{A}^\alpha\uo{\epsilon}{\alpha\mu}{\nu}\delta n^\mu\sum_{i=1}^{4}L_\nu^i\stackrel{\text{cons}}{=}0,\label{nvar1}
\end{equation}
where $\mathcal{A}^\alpha=\Gamma^\alpha+\beta\mathrm{K}^\alpha$, with $\mathcal{A}^\alpha n_\alpha=0$, denotes again the $SU(2)_n$ Ashtekar--Barbero connection \eref{ABdef}. The Gauß constraint can therefore not contribute to the time-evolution of $X^\alpha$: If the Hamiltonian vector-field of $X^\alpha$ hits $G_\alpha$, it can only generate a variation of $n^\alpha$, schematically: $\{X^\alpha,G_\alpha\}=\delta_n[G_\alpha]$, for some appropriate choice of  $\delta n^\alpha$. We have however just seen in \eref{nvar1} that $\delta_n[G_\alpha]=0$, hence: $\dot{X}^\alpha=\{H_{\text{phys}},X^\alpha\}\stackrel{\text{cons}}{=}\{\mathcal{A}^\alpha G_\alpha+NC,X^\alpha\}\stackrel{\text{cons}}{=}N\{C,X^\alpha\}$.

We are thus left to study $\{C,X^\alpha\}=\frac{1}{2}\{p_\alpha p^\alpha+\mathrm{Vol},X^\alpha\}$. Now, the second term does not contribute either: First of all, the three-volume of a tetrahedron is invariant under global $SO(3)_n$ transformation of the fluxes, because any $SO(3)_n$ transformation $L^i_\alpha\mapsto\exp(\omega)^\beta{}_\alpha L^i_\beta$, with $\ou{\omega}{\alpha}{\beta}=\ou{\epsilon}{\alpha}{\mu\beta}\omega^\mu\in\mathfrak{so}(3)_n$, preserves the cross-product $\mathrm{Vol}^2\propto\epsilon^{\alpha\beta\mu}L_\alpha^iL_\beta^jL_\mu^k$. And we have just learnt in \eref{fluxvar} that any variation of the common time-normal generates such an infinitesimal $SO(3)_n$ transformations: $\delta_n[L_\alpha^i]\stackrel{\text{cons}}{=}-\beta\uo{\epsilon}{\alpha\mu}{\beta}\delta n^\mu L^i_\beta$. The only surviving term can therefore only be the following:
\begin{equation}
\delta_n\big[\mathrm{Vol}^2\big]
\stackrel{\text{cons}}{=}\pm\frac{2}{9}\frac{\beta^3\ellp^6}{\hbar^3}\delta n_\alpha\epsilon^{\alpha\beta\mu\nu}L^i_\beta L^j_\mu L^k_\nu.
\end{equation}
Yet $\delta n^\alpha$ is purely spatial: 
$n_\alpha\delta n^\alpha=0$, and so are all the fluxes: $L_\alpha^in^\alpha=0$. In four dimensions, a quadruple of non-vanishing three-vectors can only be linearly dependent, hence:
\begin{equation}
\delta_n\big[\mathrm{Vol}^2\big]\stackrel{\text{cons}}{=}0,
\end{equation}
thus
\begin{equation}
\frac{\di}{\di t}X^\alpha=\big\{H_{\text{phys}},X^\alpha\big\}\stackrel{\text{cons}}{=}Np^\alpha.\label{Xevolv}
\end{equation}
Equations \eref{Pevolv} and \eref{Xevolv} determine the $t$-evolution of the canonical pair $(X^\alpha, p_\alpha)\equiv(X^\alpha_e, p_\alpha^e)$ along an elementary edge $e:[0,1]\ni t\mapsto e(t)\in M$. Each edge is bounded by two vertices, $v=e(0)$ and $v'=e(1)$, and each of these vertices is five-valent: There are $N$ incoming edges $v=e_1(1), \dots, e_N(1)$ and $5-N$ outgoing edges: $v=e_{N+1},\dots, e_5$. The variation of the vertex-term in the action \eref{vertxactn} imposes the boundary conditions at $v$: $\forall n=1,\dots,N;$ $n'=N+1,\dots,5: X^\alpha_{e_n}(1)=X^\alpha_{e_{n'}}(0)$, and then there is also the closure constraint for the volume-weighted normals of the tetrahedra, i.e. the equation% \eref{fourclos}: 
\begin{equation}
\sum_{n=1}^Np_\alpha^{e_n}=\sum_{n=N+1}^5p_{\alpha}^{e_{n}}.
\end{equation}
%\emph{energetic causal-set} \cite{Cortes:2013pba,Cortes:2013uka}---a system of

In other words: The collection $\{X^\alpha_e,p_\alpha^e\}_{e:\text{edges}}$ of $X$- and $p$-fields over all edges glues to form an \emph{energetic causal-set} \cite{Cortes:2013pba,Cortes:2013uka}---a system of point particles propagating in a locally flat auxiliary geometry. There is, therefore, a striking correspondence between geometry on the one side, and mechanics on the other side: Every edge $X^\alpha_e(t)$ represents a freely moving particle's worldline, %The entire simplicial complex represents a system of massive point particle scattering in a locally flat auxiliary manifold,
the tetrahedron's volume turns into the particle's rest mass, the tetrahedron's future oriented volume-weighted time-normal into the particle's four-momentum, every four-simplex represents a scattering of five such particles, and the closure constraint \eref{fourclos} for the volume-weighted time-normals is nothing but the conservation law for the total four-momentum. The $X$-variables do, however, not probe physical distances. The entire physical metric can be reconstructed from the $(\pi, \omega)$-spinors alone, which brings us to our next point---the geometric interpretation of the equations of motion in terms of twisted geometries.

%At this point, the correspondence should become obvious: The $X^\alpha$-variable evolves just as a free particle's position variable. Every tetrahedron represents a massive point particle moving in a locally flat auxiliary spacetime, the tetrahedron's volume represents the particle's rest mass, the tetrahedron's future oriented time-normals as the particle's four-momentum, the simplicial vertices as the scattering of five such particles, and finally, the closure constraint \eref{fourclos} at the vertices as the conservation law for the total four-momentum.  The $X$ -variables do, however, not probe physical distances, in fact, all the metric information on distances, angles, areas and volumes are already in the $(\pi, \omega)$-spinors.

%At this point the correspondence should become obvious: A simplicial discretization consists of $N$ tetrahedra $T_e$ glued around the four-simplices, representing $N$ particles of mass $M_e=\mathrm{Vol}(e)$. Every tetrahedron has ad future-oriented volume-weighted time-normals $p_\alpha^e=n_\alpha^e\mathrm{Vol}(e)$b$, representing the particles four-momentum $\alpha$, and at any four-simplex five bounding tetrahedra come together, such that the closure costraint \eref{fourclos} is satisfied. On the other hand, $N$ 

%succor scattering of $N$ point particles, each one of which moves along a trajectory $X_e^\alpha(t)$ in a locally flat spacetime geometry, of mass  
\subparagraph*{(iii) The emergence of twisted geometries} The next question concerns the geometric interpretation of the spinors. If we pick a point on the constraint hypersurface $\varDelta_i=G_\alpha=W_i=V_i=C=0$ and follow the Hamiltonian flow, what kind of four-dimensional geometries does this correspond to? It corresponds to a twisted geometry \cite{freidelsimotwist, twist, twistedconn}, and this can be seen as follows: First of all, any configuration on the constraint hypersurface $\varDelta_i=G_\alpha=W_i=V_i=C=0$ describes a (possibly degenerate) tetrahedron: The simplicity constraint $W_i=V_i=0$ impose that the $\mathfrak{so}(1,3)$-fluxes\footnote{The definition of the simplical fluxes involves additional holonomies mapping all Lorentz indices $\alpha,\beta,\dots$ into a common frame, here we have dropped them to simplify our notation, equation \eref{fluxdef} gives the precise definition.} $\Sigma_{\alpha\beta}[\tau_i]=\int_{\tau_i}\Sigma_{\alpha\beta}$ are simple: $\Sigma_{\alpha\beta}[\tau_i]n^\alpha=0$, hence define a plane in the local frame of reference. The Gauß constraint $G_\alpha=-\sum_{i=1}^4L_\alpha^i=0$ tells us that these planes close to form a tetrahedron, with $\Sigma_{\mu\nu}[\tau_i]=\beta\ellp^2\uo{\epsilon}{\mu\nu}{\alpha}L_\alpha^i$ linking the simplicial fluxes with the generators of $SU(2)_n$ transformations (the equation $\Sigma_{\mu\nu}[\tau_i]=\beta\ellp^2\uo{\epsilon}{\mu\nu}{\alpha}L_\alpha^i$ is a consequence of the simplicity constraints \eref{KLsimpl} and (\ref{pisigma}, \ref{LKdef})). The connection between the non-Abelian Gauß constraint on the one side, and the existence of a unique geometric tetrahedron has been well established \cite{polyhdr, bianchisommer, twist2, Barbieri:1997ks, kapovich1996} in the literature: The Minkowski theorem \cite{Minkowskitheorem} implies that any quadruple of vectors $\vec{E}_i\in\R^3$ that sum up to zero $\sum_{i=1}^4\vec{E}_i=0$, represents a (possibly degenerate) tetrahedron in $\R^3$, unique only up to rigid translations, such that: $\vec{E}_i=\mathrm{Ar}_i\vec{N}_i$, where $\mathrm{Ar}_i$ is the $i$-th triangle's area, while $\vec{N}_i$ denotes its (outwardly pointing) normal. 

The generators $L_\alpha^i\in\R^4$ are purely spatial $L_\alpha^in^\alpha=0$, hence lie in a three-dimensional real vector space $\simeq \R^3$. Thanks to the Gauß constraint, they sum up to zero. We can then employ the Minkowski theorem and reconstruct a unique tetrahedron (unique only up to rigid translations) lying in the three-dimensional subspace orthogonal to the time-normal $n^\alpha$. Every triangle $\tau_i$ of this tetrahedron has an area $\mathrm{Ar}_i=\beta\ellp^2\sqrt{-\eta^{\alpha\nu}L_\alpha^iL_\nu^i}$, while its outwardly oriented normal---with respect to the three-dimensional subspace defined by $n^\alpha$---points into the direction of $L_\alpha^i$.

Now, how does this tetrahedron evolve under the action of the physical Hamiltonian $H_{\text{phys}}$, defined as in \eref{physham}? First of all, we know that a tetrahedron is uniquely characterized by six numbers. Indeed, the length of its six bounding sides completely fixes the shape of a tetrahedron. Its volume together with the area  $\mathrm{Ar}_i=\beta\ellp^2\sqrt{-\eta^{\alpha\nu}L_\alpha^iL_\nu^i}$ of its four bounding triangles are a set of five functionally independent numbers, so we are missing yet another degree of freedom. The simplest choice is to use one of the dihedral angles:\footnote{Where again the summation convention only holds for the Lorentz indices $\alpha, \beta, \dots$; indices $i,j,k,\dots$ refer to the actual triangle $\tau_i$ under consideration, and their relative position has no geometrical meaning, e.g.: $L_\alpha^i=\eta_{\alpha\beta}L^\beta_i$.}
\begin{equation}
\cos\Theta_{ik}=
-\frac{L_\alpha^iL^\alpha_k}{\sqrt{-L_\nu^iL^\nu_i}\sqrt{-L_\nu^kL^\nu_k}}.
\end{equation}
The squared volume functional \eref{voldef} is a polynomial in the $SU(2)_n$ generators $\mathrm{Vol}^2=\frac{2}{9}\epsilon^{\alpha\beta\mu}L_\alpha^iL_\beta^jL_\mu^k$, and commutes, therefore, with the $SU(2)_n$ invariants $\propto L_\alpha L^\alpha$. The area of a triangle is just the square root of this invariant: $\mathrm{Ar}_i=\beta\ellp^2\sqrt{-L_\alpha^iL^\alpha_i}$, hence:
\begin{equation}
\big\{\mathrm{Vol}^2,\mathrm{Ar}_i\big\}=0,\quad\text{and trivially:}\quad \big\{\mathrm{Vol}^2,\mathrm{Vol}^2\big\}=0.
\end{equation}
We thus see, that the Hamiltonian flow of the volume functional preserves both the four areas and the volume of the tetrahedron. Yet it does not preserve the remaining sixth degree of freedom:  
\begin{equation}
\big\{\mathrm{Vol}^2,\cos\Theta_{kl}\big\}
\stackrel{\text{in general}}{\neq}0.
\end{equation}
Bianchi and Haggard \cite{bianchisommer} have given a very detailed analysis of this Hamiltonian flow in the space of all possible geometric tetrahedra: If we keep the four areas $\mathrm{Ar}_i$ fixed, the two remaining degrees of freedom form a phase space. This phase space has the topology of a two-sphere, and the orbits of the volume functional are closed lines winding once around the two poles of this two-sphere. Different points on a given orbit represent tetrahedra of equal areas and volume, yet different shape.

The physical Hamiltonian is a sum over constraints. The Gauß law $G_\alpha=-\sum_{0=1}^4L_\alpha^i$ generates $SU(2)_n$ transformations, hence commutes with the $SU(2)_n$ invariants $\cos\Theta_{ij}$, $\mathrm{Vol}$ and $\mathrm{Ar}_i$. Both the area-matching constraints $\varDelta=\utilde{E}_A-E$ and the simplicity constraint $V_i=\I/(\beta+\I)E+\CC$ are a linear combination of the $SL(2,\C)$ invariant contraction $E=\pi_A\omega^a$ of the spinors, hence trivially commute with the $SU(2)_n$ generators $L_\alpha^i$ as well, thus:
\begin{equation}
\big\{V_i,L_\alpha^j\big\}=\big\{\varDelta_i,L_\alpha^j\big\}=0.
\end{equation}
The mass shell condition $C=\frac{1}{2}(p_\alpha p^\alpha+\mathrm{Vol}^2)$ is therefore the only term that determines the evolution of the tetrahedron's shape:
\begin{equation}
\frac{\di }{\di t}\cos\Theta_{ik}
=\big\{H_{\text{phys}},\cos\Theta_{ik}\big\}=
N\big\{C,\cos\Theta_{ik}\big\}=
\frac{N}{2}\big\{\mathrm{Vol}^2,\cos\Theta_{ik}\big\}\stackrel{\text{in general}}{\neq}0,
\end{equation}
while
\begin{equation}
\frac{\di }{\di t}\mathrm{Ar}_i
=
\big\{H_{\mathrm{phys}},\mathrm{Ar}_i\}=0,\qquad
\frac{\di }{\di t}\mathrm{Vol}=
\big\{H_{\mathrm{phys}},\mathrm{Vol}\}=0.
\end{equation}

In other words, the Hamiltonian flow $\{H_{\text{phys}},\cdot\}$ generates a shear: Both the tetrahedron's volume and the area of its four bounding triangles are preserved; only the tetrahedron's shape changes as we move forward in $t$. Any triangle $\tau_f$ in the simplicial complex has a well defined area, but its shape is a continuous function over the boundary $\partial f$ of the dual face. The resulting geometry is therefore very different from a Regge discretization. In Regge calculus the lengths of the one-dimensional bones of the triangulation are the elementary configuration variables, all tetrahedra consistently glue together, and every triangle has a unique shape. In our case the situation is different, the three sides of a triangle do not have a unique length, and the resulting geometry is a \emph{twisted geometry} \cite{freidelsimotwist, twist, twistedconn, contphas, Freidel:2014aa}. A twisted geometry (in four dimensions) is an (oriented) four-dimensional simplicial complex with a locally flat Lorentz metric in each four-simplex, along with the following two glueing conditions:
 The first condition is that the metrics in any two neighboring four-simplices agree on the volume of the interjacent tetrahedron. The second condition is that every triangle has a unique area whether it is measured with the metric of one adjacent four-simplex or the other.
% 
% 
% for any collection of four-simplices sharing a triangle, the triangle's area is the same areal does not depend on whether we measure it from the metric of one adjacent four simplex or the other.
% 
% four-simplex sharing a triangle the triangle has a unique area we measure the area with the metric of one adjacent four-simplex or the other.
% 
% 
% for any two tetrahedra sharing a triangle the triangle's area computed is the same from the metric on either side. It does not matter whether we compute it from the metric on one side or the other, we would always get the same result. 
%

In our case, there are no unique length variables either---a collection of four-simplices may not agree on the shape of an interjacent triangle. This is one of the key novelties of our formalism, and it provides the first concrete mechanism that generates a twisted geometry from the variation of an action.%: An edge connects two adjacent four-simplices, and the physical Hamiltonian \eref{physham} that governs the $t$-evolution along this edge does not preserve the shape of the dual tetrahedron. 
%Our formalism provides the first concrete example of a mechanism that dynamically generates twisted geometries from the variation of an action.
% Each tetrahedron belongs to two four-simplices. Given two tetrahedra sharing a triangle there are therefore no more than $2\times 2=4$ different metrics available to compute the triangle's area. 

\subparagraph*{(iv) gauge symmetries} Looking back at the constraint algebra (\ref{fclass}, \ref{sclass}), we can now immediately identify the local gauge symmetries of the model: %The physical Hamiltonian \eref{physham} $H_{\text{phys}}=\mathcal{A}^\alpha G_\alpha+NC+\sum_{i=1}^4(\zeta^i\varDelta^i+\bar{\zeta}^i\bar{\varDelta}_i+\lambda^iV_i)$ is a sum over all first-class constraints of the theory, and 
%There are four independent local gauge symmetries: %, and its Hamiltonian flow thus annihilates all constraints of the theory.
The Gauß constraint $G_\alpha=-\sum_{i=1}^4L_\alpha^i$ is a sum over the $SU(2)_n$ generators $L_\alpha^i=-\ou{\tau}{AB}{\alpha}\omega^i_A\pi^i_B+\CC$, hence generates simultaneous $SU(2)_n$ transformations of all \qq{untilded spinors}: $(\pi^i_A,\omega^i_A)\mapsto (\ou{U}{B}{A}\pi^i_B,\ou{U}{B}{A}\omega^i_B)$ with $U\in SU(2)_n$. The constraints $\varDelta_i$ and $V_i$, on the other hand, generate overall rescalings of the spinors:  The exponential of the Hamiltonian vector field of $\varDelta_i$ yields the map: $(\pi^i,\omega^i,\utilde{\pi}^i,\utilde{\omega}^i)\mapsto(\E^{\zeta}\pi^i,\E^{-\zeta}\omega^i,\E^{\zeta}\utilde{\pi}^i,\E^{-\zeta}\utilde{\omega}^i)$, with $\zeta\in\C$. The simplicity constraints $V_i$ only transform the corresponding $(\pi^i,\omega^i)$-spinors and generate the twisted transformations $\omega^i\mapsto\exp({\frac{\I}{\beta+\I}\lambda}){\omega^i}$ and $\pi^i\mapsto\exp({-\frac{\I}{\beta+\I}\lambda}){\pi}^i$, with $\lambda\in\R$. Finally, we also have the mass shell condition $C=\frac{1}{2}p_\alpha p^\alpha+\mathrm{Vol}^2$. As we have just seen above, its Hamiltonian vector field generates shear transformations of the elementary tetrahedra. This summarizes the \emph{local} gauge symmetries of the model. Yet there may be additional symmetries. In fact, we can immediately give an example for a non-local gauge symmetry: The dynamics of the tilde-spinors $\tilde{\pi}^A_f:\partial f\rightarrow \C^2$ and $\tilde{\omega}^A_f:\partial f\rightarrow \C^2$ is pure gauge. If $\utilde{\pi}^A_f(t)$ and $\tilde{\omega}^A_f(t)$ are a solution of the Hamilton equations \eref{tildedyn} subject to both the area matching constraint $\varDelta_f=0$ and the periodic boundary conditions $\utilde{\pi}^A_f(0)=\utilde{\pi}^A_f(1)$ and $\utilde{\omega}^A_f(0)=\utilde{\omega}^A_f(1)$, then we can always pick any $g\in SL(2,\C)$ and construct the transformed fields $\ou{g}{A}{B}\utilde{\omega}^B(t)$ and $\ou{g}{A}{B}\utilde{\pi}^B(t)$, which are again a solution of the equations of motions---the transformed spinors still solve the area matching constraint, they solve the periodic boundary conditions and also solve the evolution equations \eref{tildedyn}.

\subsection{Curvature and deficit angles}\noindent We will now close this paper with an analysis of the deficit angle around a triangle, which is a measure for the curvature in the dual face.
Consider an edge $e:[0,1]\rightarrow\partial f\subset M$ in the boundary of a face $f$, connecting two vertices $v_o$ and $v_1$ of the triangulation: $e(0)=v_o$ and $e(1)=v_1$. The edge carries a time-normal $n^\alpha$, and we thus have two time normals $n^\alpha|_{e(1)}$ and $n^\alpha|_{e(0)}$ sitting at the two boundary points $e(0)=v_o$ and $e(1)=v_1$ of the edge. What is the Lorentz angle between the two? Since the normals belong to different points in the manifold, their inner product has a geometrical interpretation only if we map them into a common frame. Let us choose the frame at the center of the face $f$. We have already introduced in (\ref{linkhol}, \ref{holpar}) the $SL(2,\C)$ parallel transport $h(t)=\mathrm{Pexp}(-\int_{\gamma_t}A)$ along the connecting link $\gamma_t$. This holonomy is a map from the Lorentz fibre over the boundary $e(t)\in\partial f$ of $f$ into the fibre over the center $c=\tau_f\cap f$ of the face (where $\tau_f$ is the triangle dual to $f$). In the spinorial representation this parallel transport becomes:
\begin{equation}
\ou{h}{A}{B}(t)=\frac{\utilde{\omega}^A(t)\pi_B(t')-\utilde{\pi}^A(t)\omega_B(t)}{E(t)},\label{norm1}
\end{equation}
where $E(t)=\pi_A(t)\omega^A(t)=\utilde{\pi}_A(t)\utilde{\omega}^A(t)$. We can now map the time normal $n^\alpha_e\equiv n^\alpha$ into the frame at the center of the face:
\begin{equation}
\utilde{n}^{AA'}(t)=\ou{h}{A}{B}(t)\ou{\bar h}{A'}{B'}(t)n^{BB'}(t),\label{norm2}
\end{equation}
where $n^\alpha(t_o)=n^\alpha(t_1)$ thanks to the evolution equations \eref{Pevolv} for the volume weighted time-normal $n^\alpha\propto p^\alpha$. For any two values $0\leq t_o<t_1\leq 1$ of $t$, corresponding to two consecutive points $e(t_o)$ and $e(t_1)$ in the boundary of $f$, we can now define the Lorentz angle in between:
\begin{equation}
\cosh\Xi(t_o,t_1)=-\utilde{n}_\alpha(t_o)\utilde{n}^\alpha(t_1)=
-\utilde{n}_{AA'}(t_o)\utilde{n}^{AA'}(t_1).\label{Xidef}
\end{equation}
We now take this expression, and insert the holonomies in the spinorial representation \eref{norm1}. Employing the simplicity constraints $W=n^{AA'}\pi_A\bar{\omega}_{A'}=0=\I/(\beta+\I)\pi_A\omega^A+\CC=V=0$ together with the area-matching constraint $\varDelta=0\Leftrightarrow \pi_A\omega^A=\utilde{\pi}_A\utilde{\omega}^A$ eventually leads us to the following expression:
\begin{equation}
\cosh\Xi(t_o,t_1)=\frac{1}{2}\bigg[\E^{\int_{t_o}^{t_1}\di t\left(\zeta(t)+\bar{\zeta}(t)\right)}\frac{\|\omega(t_1)\|^2_n}{\|\omega(t_o)\|^2_n}+
\E^{-\int_{t_o}^{t_1}\di t\left(\zeta(t)+\bar{\zeta}(t)\right)}\frac{\|\omega(t_o)\|^2_n}{\|\omega(t_1)\|^2_n}\bigg].
\end{equation}
The spinors only appear in the fraction $\|\omega(t_1)\|^2_n/\|\omega(t_o)\|^2_n$ of the $SU(2)_n$ norm $\|\omega\|^2_n=\sigma_{AA'\alpha}n^\alpha\omega^A{\bar\omega}^{A'}$. We employ the  Hamilton equations $\frac{\di}{\di t}\|\omega\|^2_n=\{H_{\text{phys}}, \|\omega\|^2_n\}$ and compute this fraction explicitly: The Hamiltonian is a sum over constraints, but only some of them contribute to the evolution of $\|\omega\|^2_n$: The Gauß constraint generates $SU(2)_n$ transformations, hence trivially Poisson-commutes with the $SU(2)_n$ norm $\|\omega\|^2_n$. Gauge invariance also implies that the mass-shell condition $C=\frac{1}{2}(p_\alpha p^\alpha+\mathrm{Vol}^2)$ does not contribute: $\mathrm{Vol}^2$ is a polynomial (see again \eref{voldef}) of three $SU(2)_n$ generators $L^i_\alpha=-\ou{\tau}{AB}{\alpha}\omega_A^i\pi_B^i+\CC$ $i=1,2,3$, each of which Poisson-commutes with $\|\omega\|^2_n$. Finally $\{p_\alpha p^\alpha,\|\omega\|^2_n\}=0$ because $p^\alpha$ already Poisson-commutes with both $n^\alpha$ and $\omega^A$ itself. Going back to the definition of the physical Hamiltonian \eref{physham}, the only relevant terms, are, therefore: 
\begin{align}\nonumber
\frac{\di}{\di t}\|\omega(t)\|^2_n&=
\Big\{\mathcal{A}^\alpha G_\alpha+\zeta\varDelta+\bar{\zeta}\bar\varDelta+\lambda V+\frac{N}{2}\left(p_\alpha p^\alpha+\mathrm{Vol}^2\right),\|\omega\|^2_n\Big\}\Big|_{e(t)}=\\
&=\Big\{\zeta\varDelta+\bar{\zeta}\bar\varDelta+\lambda V,\|\omega\|^2_n\Big\}\Big|_{e(t)}=
-\Big(\zeta(t)+\bar\zeta(t)-\frac{2}{\beta^2+1}\lambda(t)\Big)\|\omega(t)\|^2_n.
\end{align}
We can immediately integrate this equation to find:
\begin{equation}
\|\omega(t_1)\|^2_n=\E^{-\int_{t_o}^{t_1}\di t\left(\zeta(t)+\bar\zeta(t)-\frac{2}{\beta^2+1}\lambda(t)\right)}\|\omega(t_o)\|^2_n.\label{normevolv}
\end{equation}
Hence
\begin{equation}
\cosh\Xi(t_o,t_1)=\frac{1}{2}\left[\E^{\frac{2}{\beta^2+1}\int_{t_o}^{t_1}\di t\lambda(t)}+
\E^{-\frac{2}{\beta^2+1}\int_{t_o}^{t_1}\di t\lambda(t)}\right].\label{bangle}
\end{equation}
This equation defines the boost angle only up to an overall sign. We remove this ambiguity by setting:
\begin{equation}
\Xi(t_o,t_1):=\frac{2}{\beta^2+1}\int_{t_o}^{t_1}\!\!\!\di t\,\lambda(t).\label{Xispin}
\end{equation}
So far, we have just studied the boost angle between to consecutive points $e(t_o)$ and $e(t_1)$ in some given edge $e$. Let us now see how these angles probe a Regge-like deficit angle once we sum them up around a face. Let us first define the boost angle between two adjacent tetrahedra belonging to the same four-simplex. Two adjacent tetrahedra are dual to a pair of edges $e$ and $e'$ meeting at a vertex $v$, with the vertex representing the four-simplex containing the two tetrahedra $T_e$ and $T_{e'}$. The two tetrahedra have future oriented time normals $n^\alpha_e$ and $n^\alpha_{e'}$ respectively, with an angle $\Xi_{vf}$ in between:
\begin{equation}
\cosh\Xi_{vf}=-\eta^{\mu\nu}n_\mu^en_\nu^{e'},\quad \text{with:}\; e\cap e'=v,\;\text{and:}\; e,\,e'\subset\partial f,\label{Xivertex}
\end{equation}
where $f$ denotes the face, that is bounded by both $e$ and $e'$: $e\subset \partial f$ and $e'\subset \partial f$. To remove the sign ambiguity in the definition of $\Xi_{vf}$ we proceed as follows: First of all, we can always choose the orientation of the two edges $e:[0,1]\rightarrow\partial f$ and $e':[0,1]\rightarrow\partial f$ such that they have the same orientation as $\partial f$. We can then also assume that $e$ comes \qq{before} $e'$, i.e.: $e(1)=e'(0)$. The arguments that have led us from the definition of the Lorentz angle $\Xi(t_0,t_1)$ \eref{Xidef} to its integral representation \eref{Xispin} also apply to the boost angle $\Xi_{vf}$ at the vertex. Repeating these arguments eventually brings us to the following expression:
\begin{equation}
\Xi_{vf}=
\log\frac{\|\omega_f\|^2_{n_e\phantom{'}}}{\|\omega_f\|^2_{n_{e'}}}\bigg|_v,\quad\text{such that:}\;e(1)=e'(0)=v,\;\text{and:}\; e, e'\subset\partial f\label{Xidef2}
\end{equation}
where $\|\omega_f\|^2_{n_e}|_v=n_e^\alpha\sigma_{AA'\alpha} \omega^A_f|_v\,\bar{\omega}^{A'}_f|_v$ denotes the $SU(2)_n$ norm with respect to the time-normal over the edge $e\subset\partial f$, while $\omega_f^A|_v$ is the value of the spinor $\omega^A_f:\partial f\rightarrow \C^2$  at the vertex $v\in e\subset\partial f$. We have thus implicitly removed the sign ambiguity $\Xi_{vf}\rightarrow -\Xi_{vf}$ in equation \eref{Xivertex} and gave an unambiguous definition for $\Xi_{vf}$ compatible with $\cosh \Xi_{vf}=-n^\alpha_{e}n_\alpha^{e'}$. 

What is remarkable is that the boost angles at the vertices, i.e. $\Xi_{vf}$, and the boost angles along the edges, e.g. $\Xi(t_o,t_1)$ as defined in \eref{Xispin}, are not independent. This can be seen as follows: Consider first the sum over all boost-angles between adjacent tetrahedra in a face: 
\begin{equation}
\Xi_f:=\!\!\!{\sum_{\text{$v$:\,vertices\,in\,$f$}}}\!\!\!\Xi_{vf}=\!\!\!{\sum_{\text{$e$:\,edges\,in\,$f$}}}\!\!\!\log\frac{\|\omega_f\|^2_{n_e}\big|_{e(1)}}{\|\omega_f\|^2_{n_{e}}\big|_{e(0)}},
\end{equation}
where we have implicitly assumed that the orientation of the edges $e\subset \partial f$ matches the induced orientation of $\partial f$ (otherwise additional sign factors would be necessary). Inserting the general solution for $\|\omega(t)\|^2_n$ as derived in \eref{normevolv}, we thus get:
\begin{equation}
\Xi_f=\!\!\!{\sum_{\text{$v$:\,vertices\,in\,$f$}}}\!\!\!\Xi_{vf}=-\int_{\partial f}\left(\zeta_f+\bar\zeta_f-\frac{2}{\beta^2+1}\lambda_f\right)=\frac{2}{\beta^2+1}\int_{\partial f}\lambda_f,\label{Xidef3}
\end{equation}
where the first two terms in the parenthesis cancel another thanks to equations (\ref{BFflata}, \ref{windnumb}): $\zeta_f$ cancels $\bar\zeta_f$ because the integral of the Lagrange multiplier $\zeta_f$ over the boundary of the face is purely imaginary: $\int_{\partial f}\zeta_f=2\pi\I\,n_f$. There is therefore a crucial difference between the two
 boost-angles %\footnote{It seems to me that the failure to make this distinction has created a substantial confusion in the loop quantum gravity literature, and that the flatness issue that has been pointed out by Bonzom, Hellmann and Kaminski \cite{Bonzom:2009hw,Hellmann:2013gva} concerns the integral $\int_{\partial f}\zeta_f$ and not the deficit angle $\Xi_f$ itself.}  
$\Xi_f$ and $\int_{\partial f}\zeta_f=2\pi\I\, n_f$: The former (i.e. $\Xi_f$) probes the curvature in a face, while the latter (i.e. $\int_{\partial f}\zeta_f$) vanishes modulo $2\pi\I$.

%\section{Applications}\label{secIII}
%\subparagraph{The Minkowski solution}
%\subparagraph{Curvature}
%\subparagraph{Torsion}
%\newpage
\section{Conclusion}\noindent
\subparagraph*{Summary} We have split our presentation into two halves. The first half gave the derivation of the action. We started with the topological $SO(1,3)$ BF action in four dimensions. A simplicial decomposition of the four-dimensional manifold brought us to the discretized action, and we saw that this discretized action can be written as a one-dimensional integral over the simplicial edges---as an action over a one-dimensional branched manifold. The spinorial representation turned the action into a sum over three terms: There is the symplectic potential, the area-matching constraint reducing the spinors to holonomy-flux variables and the Gauß constraint generating the local $SO(1,3)$ gauge symmetries of the theory. We went further and studied the discretized simplicity constraints. In the continuum, the simplicity constraints reduce the topological BF action to the Holst action for general relativity, and we expect that this is also true in the discrete theory. We then studied the discretized simplicity constraints in the spinorial representation and added them to our one-dimensional action. This introduced an additional element to the theory: The volume-weighted time-normals of the elementary tetrahedra. We argued that a consistent theory is possible only if we also treat these time normals as dynamical variables in the action. At each simplicial vertex these volume-weighted normals sum up to zero---this representing the geometricity of the four-simplex itself. We argued that this closure constraint represents a momentum conservation law for a system of particles scattering in a locally flat auxiliary spacetime, and we made an explicit proposal for an action realizing this idea. Each of these particles corresponds to an elementary tetrahedron in the simplicial complex, with their mass representing the three-volume of the elementary tetrahedra, and each interaction vertex representing a four-simplex in the discretization.

The second half of the paper studied the dynamics of the theory as derived from the action. Let us say it clearly: We have not shown that the equations of motion for the discretized theory would correspond to some version of the Einstein equations on a simplicial lattice. Nevertheless, we do have some definite results: First of all, there is a Hamiltonian formulation for the discretized theory, there is a phase space, constraints and a Hamiltonian. The Hamiltonian generates the $t$-evolution along the elementary edges of the discretization, and preserves both the first- and second class constraints. There are no secondary constraints. Next, we showed that the solutions of the equations of motion have a geometric interpretation in terms of \emph{twisted geometries}. Twisted geometries are piecewise flat geometries generalizing Regge geometries: In Regge calculus the edge lengths are the fundamental configuration variables, in twisted geometries there is no unique notion of length: Every tetrahedron has a unique volume, and every  triangle has a unique area, but the length of a segment depends on whether we compute it from the flat metric in one simplex or the other.
Finally we gave an argument why the model has curvature. Going around a triangle we pick up a deficit angle, which is a measure for the curvature in the dual plane. We showed that this deficit angle will generically not vanish, in fact it is given by the integral of the Lagrange multiplier $\lambda$ (imposing the $V$-component \eref{spinsimplb} of the linear simplicity constraints) over the loop bounding the face. This should not come as a surprise: If we take the Pleba\'{n}ski action $S=\int_M(\Sigma_{\alpha\beta}\wedge F^{\alpha\beta}+\tfrac{1}{4}C^{\alpha\beta\mu\nu}\Sigma_{\alpha\beta}\wedge\Sigma_{\mu\nu})$, then we will also get that the curvature tensor is linear in the Lagrange multiplier enforcing the simplicity constraints: $F_{\alpha\beta}=\frac{1}{2}C_{\alpha\beta\mu\nu}\Sigma^{\mu\nu}$, and equation \eref{Xidef3} is the manifestation of this mechanism in the discrete theory.
\subparagraph*{The relevance of the model} The action \eref{spinactn} describes a system of finitely many degrees of freedom propagating and interacting along the simplicial edges. The system has a phase space, local gauge symmetries and a Hamiltonian. What happens if we quantize this model? Do we get yet another proposal for a theory of quantum gravity? Recent results \cite{CortezLee,Wieland:2014vta,hamspinfoam,twistintegrals} point into a more promising direction and suggest a convergence of ideas: The finite-dimensional phase space can be trivially quantized. The constraints of the theory glue the quantum states over the individual edges so as to form a Hilbert space over the entire boundary of the underlying simplical manifold. The boundary states represent projected spin-network functions \cite{sufromcov,liftng} in the kinematical Hilbert space of loop quantum gravity. It is clear what should be done next: For any fixed boundary data we should define a path integral over the field configurations along the edges in the bulk. At this point, many details remain open, and we have only finished this construction for the corresponding model in three-dimensions \cite{Wieland:2014vta}, yet we do know, that whatever the mathematical details of the resulting amplitudes will be, they will define a version of spinfoam gravity \cite{reisenberger}. Finally, there is the motion of the volume-weighted time normals, which endow the entire simplicial complex with a flow of conserved energy-momentum. 
As shown by Cortês and Smolin in a related paper \cite{CortezLee}, these \emph{momentum-variables} introduce a causal structure, and allow us to view the simplicial complex as an energetic causal set \cite{Cortes:2013uka,Cortes:2013pba}---a generalization of causal sets carrying a local flow of energy-momentum between causally related events.

\subparagraph*{Acknowledgements} It is my pleasure to thank Abhay Ashtekar, Eugenio Bianchi, Marina Cortês, Marc Geiller and Lee Smolin for helpful comments and discussions. I gratefully acknowledge support from the Institute for Gravity and the Cosmos, the National Science Foundation through grant PHY-12-05388 and the Eberly research funds of The Pennsylvania State University.

\providecommand{\href}[2]{#2}\begingroup\raggedright\endgroup

%\appendix
%\bibliography{/Users/wmwieland/Desktop/Projekte/Bibliographie/LQG-1.bib}
%\bibliographystyle{/Users/wmwieland/Desktop/Projekte/Bibliographie/disstyle.bst}
%Consider a triangulated four-dimensional manifold $\mathcal{M}$. There is no metrical structure yet, all that we have are:
%\begin{center}
%\begin{tabular}{r l r l}\hline
%  &{\bf a 4d simplicial complex,} & & \bf and its dual.} \\\hline
%  &four-simplices &  $(v)$ &vertices \\
%  $(T)$&tetrahedra & $(e)$ & edges  \\
%  $(\tau)$&triangles & $(f)$ & faces \\
%  $(b)$&sides  & &  dual polytopes \\
%  &corners & & bubbles  \\\hline
%\end{tabular}
%\end{center}

%\begin{center}
%\begin{tabular}{r l r l}\hline
%  &{\bf a 3d simplicial complex,} & & {\bf and its dual.} \\\hline
%  $(T)$&tetrahedra & $(e)$ & nodes  \\
%  $(\tau)$&triangles & $(f)$ & links \\
%  $(b)$&sides  & & plane  \\
%  &corners & & dual corner \\\hline
%\end{tabular}
%\end{center}

%\begin{center}
%\begin{tabular}{r l r l}\hline
% &{\bf simplicial complex} & & {\bf scattering interpretation} \\\hline
%  & spinfoam vertices &  & interaction vertices \\
%  & tetrahedra &  & particles \\
%  & volume weighted normals &  & particle momenta  \\
%  & spinors  & &  internal colour DOFs \\\hline
%\end{tabular}
%\end{center}

\end{document}